\documentclass[useAMS,usenatbib]{mn2e}

\usepackage{graphicx}
\usepackage{amsmath}
\usepackage{amssymb}
\usepackage{color}
\usepackage{gensymb}
\usepackage{subfigure}

\title[High velocity stars from close interaction of a globular cluster and a super massive black hole]{High velocity stars from close interaction of a globular cluster and a super massive black hole}
\author[R. Capuzzo-Dolcetta and G. Fragione]{R. Capuzzo-Dolcetta$^{1}$\thanks{E-mail:
roberto.capuzzodolcetta@uniroma1.it} and G. Fragione$^{1}$\thanks{E-mail: giacomo.fragione@uniroma1.it}\\
$^{1}$Dep. of Physics, Sapienza, Univ. of Roma, P.le A. Moro 2, Roma, Italy}
\begin{document}



\maketitle

\begin{abstract}
Observations show the presence, in the halo of our Galaxy, of stars moving at velocities so high to require an acceleration mechanism involving the presence of a massive central black hole. Thus, in the frame of a galaxy hosting a supermassive black hole ($10^8$ M$_\odot$) we investigated a mechanism for the production of high velocity stars, which was suggested by the results of $N$-body simulations of the close interaction between a massive, orbitally decayed, globular cluster and the super massive black hole. The high velocity acquired by some stars of the cluster comes from the transfer of gravitational binding energy into kinetic energy of the escaping star originally orbiting around the cluster.
After the close interaction with the massive black hole, stars could reach a velocity sufficient to travel in the halo and even overcome the galactic gravitational well, while some of them are just stripped from the globular cluster and start orbiting on precessing loops around the galactic centre.
\end{abstract}


\label{firstpage}

\section{Introduction}
The existence of high velocity stars in the Galactic halo is an ascertained feature. Some of them have speed sufficient to escape the Galaxy gravitational potential. They may have gained such high velocities thanks to different physical mechanisms, as three-body interactions among binary systems in star clusters or with the massive black hole in the Galactic centre. High velocity stars can be divided in two different categories, i.e. runaway stars and hypervelocity stars.

Runaway stars, historically defined in the context of O and B stars \citep{hum47}, are Galactic halo stars with peculiar motions higher than $40$ km s$^{-1}$ (although the definition of  runaway star is not univocal). Young massive stars are not expected to be observed in the halo far from star-forming regions, since special conditions, as the presence of molecular clouds with dense cores, are required to form such stars. Therefore, they are thought to be born not in the halo, but rather to have travelled far from their birthplace. There are two proposed mechanisms for the production of runaway stars: supernova ejections and dynamical ejections \citep{sil11}.

In the supernova ejection mechanism \citep{bla61,por00} a runaway star is supposed to have origin in a binary system when its companion explodes as a supernova. The maximum possible ejection velocity is given by the sum of the orbital velocity of the progenitor binary and of the supernova kick velocity. Additional effects may come from asymmetric explosions \citep{sch06,prz08}, but, in any case, runaways velocities are below the Galactic escape velocity. In the dynamical ejection mechanism \citep*{pov67} the runaway star derives from a three- or four-body interaction. For example, if a binary system interacts with a massive star, one member of the binary could be captured by the massive star, while the other star may be ejected with high velocity \citep*{leo90,gvr09,gva09,gva11,per12}. In this case, the maximum possible ejection velocity is the escape velocity of the most massive star. Observations show that both the ejection mechanisms operate in nature \citep{hoo01}.

Hypervelocity stars (HVS) are stars escaping the host Galaxy. \citet{hil88} was the first to predict theoretically their existence as a consequence of interactions with a massive Black Hole (BH) in the Galactic Centre, while \citet{brw05} serendipitously discovered the first HVS in the outer stellar halo, a B-type star moving over twice the Galactic escape velocity. The most recent HVS Survey is the MMT (Multiple Mirror Telescope) survey, a spectroscopic survey of stars within the range of colours of $2.5-4$ M$_{\odot}$ late B-type stars \citep{brw14}. The MMT observational approach is justified by that such stars should not exist at faint magnitudes in the outer halo unless they were ejected until those distances. Actually, such stars have relatively short lifetimes and should originate in a region of on-going star formation. The MMT survey revealed $21$ hypervelocity stars ejected from the Milky Way at distances between $50$ and $120$ kpc. However, the MMT survey is able to measure only the component along the line of sight of the velocity vector. To compute the other velocity components, a measure of proper motions is needed. This is in most of the cases not possible (but for some special cases \citep{brw10}), because HVSs found by MMT Survey are very distant so that their proper motions are too small ($\le 1$  mas yr$^{-1}$) to be measured with ground-based telescopes. Moreover, as said above, the MMT Survey is biased to the observation of stars within the colour range of $2.5-4$ M$_{\odot}$ late B-type stars, although an observational effort to find an older population of HVS \citep*{brw09, kol09, kol10} has been done, unsuccessfully, in the last years.

Hills' mechanism involves the tidal breakup of a binary passing close to a massive BH. This mechanism was analyzed by other authors in the attempt to shed light on the properties of the stars that are accelerated in such a way \citep*{yut03, gua05, brm06, sar09, kob12, ros14}. The tidal breakup of a binary could lead also to a population of stars orbiting in the inner regions of the Galaxy around the central BH, the so-called S stars \citep*{gou03, gin06, per07}. Since the Hills' prediction, a lot of other mechanisms have been proposed in the literature to explain the production of high velocity and hypervelocity stars, which involve different astrophysical frameworks and phenomena \citep{tuf09}:

\begin{itemize}
\item the interaction of a Super Massive Black Hole Binary (SMBHB) with a single star in the nucleus of the host galaxy. In this way stars can be accelerated at velocities high enough to escape the local gravitational potential \citep*{gua05, bau06, ses06};
\item some stars can come from another (nearby) galaxy with a high velocity relative to the present galactic environment \citep*{gua07, bon08, she08, per09, brw10};
\item the close encounter of a hard massive binary star and a single massive star could produce stars with velocities larger than the local escape velocity \citep*{gva09};
\item a Supernova explosion in a close binary can give a kick to the companion to accelerate it to a very high velocity \citep{por00, zub13}.
\end{itemize}

An interesting point is that, as shown by \citet{han07} and \citet{lop08}, HVS in our Galaxy have both slow and rapid rotations, suggesting different acceleration mechanisms. Note indeed that HVSs originated by a binary disrupted by a Black Hole are not expected to be fast rotators and so this origin is unlikely for fast rotating HVSs. Therefore, since HVS production mechanisms should involve different astrophysical frameworks and phenomena, it would be possible to infer information about different pieces of physics, as that of the three-body interaction, the physics of the region near massive BHs \citep*{gou03, ses07, oll08} as well as the physics of Supernovae. Moreover, the study of the proper motions of such fast moving stars can improve the knowledge of the Galaxy gravitational potential shape, of its Dark Matter component \citep{gne05, yum07} and, in line of principle, may lead to useful information also for cosmology \citep{loe11}.

Observations of high velocity and hypervelocity objects have been limited to high-mass, early-type stars due to obvious observational bias. Observers have started investigating low-mass high velocity stars only recently \citep*{pal14,zho14,li15,vic15}, some of which are low-mass HVS candidates. The European ESA satellite GAIA\footnote{http://www.cosmos.esa.int/web/gaia} is expected to measure proper motions with a precision of $0.1$ mas yr$^{-1}$ and so will be able to provide for a larger and less biased sample. Furthermore, \textit{Gaia} is expected to find $\sim 100$ HVSs in a sample of $\sim 10^9$ stars.

The aim of this paper is to investigate another mechanism of production of high velocity stars, which involve a Globular Cluster (GC) that during its orbit has the chance to pass close to an SMBH in the center of its host galaxy. This chance is increased by the orbital decay suffered by massive clusters moving a dense galactic environment, which makes significant the dynamical friction braking exerted by the stars of the galaxy \citep{cap93,cap08,ant12}. For these test cases, we assumed $M_{BH}=10^8$ M$_\odot$ with the scope of identify at better the underlying physical mechanism.

The paper is organized this way: in Sect. \ref{sec:meth} we outline and describe our approach to the study of the consequences of the GC-SMBH interaction; in Sect. \ref{sec:res} the results are presented  and discussed; in Sect. \ref{sec:con} we draw the conclusions. Significant details are given in the Appendix.

\section{Method}
\label{sec:meth}
Our scattering experiments refer to the interaction of three different bodies: a super massive black hole (SMBH), a globular cluster (GC) and a star. In our simulations the SMBH sits initially in the origin of the reference frame, while the GC follows an elliptical orbit at a relatively close distance around it. The assumption of close distance to the BH is motivated by that the globular cluster is supposed orbitally decayed by dynamical friction braking, as discussed in Arca Sedda et al. (in preparation). 
Given the BH influence radius as that within which the BH potential dominates

\begin{equation}\label{eqn:infrad}
r_{inf} = \frac{GM_{BH}}{\sigma^2},
\end{equation}

where $\sigma$ is the stars velocity dispersion in the central galactic region, we can treat the dynamics of the star interacting with the environment as a 3-body (GC, SMBH and test star around the GC) problem whenever the relevant fly-by passage occurs within the distance $r_{inf}$ from the SMBH. In the cases studied in this paper (Sect. \ref{sec:res}), $r_{inf} \lesssim 12.5$ pc, which means that the choice we make in this paper of 10 pc as radius of the GC circular reference orbit, with the values of pericenters distances (around which the scattering is effective) given in Table 1, is fully compatible with the neglect of the smooth external field.

Actually, neglecting the stellar background potential, the  mechanical energy (per unit mass) of the GC on a circular orbit of radius $r_c$ is

\begin{equation}\label{eqn:ener}
E_{c}\equiv \frac{1}{2}v_c^2-\frac{GM_{BH}}{r_c}=-\frac{1}{2}\frac{GM_{BH}}{r_c},
\end{equation}

given that the circular velocity is $v_c=(GM_{BH}/r_c)^{1/2}$.
Consequently, taking into account that the angular momentum per unit mass of the GC on the circular orbit around the BH is  $L_c=\sqrt{GM_{BH}r_c}$, the pericenter ($r_-$) and apocenter ($r_+$) distances of the GC on orbits of same energy ($E_c$) but different angular momentum $0\leq L\leq L_c$ are given by 

\begin{equation}
r_{\pm} = r_c\left(1\pm \sqrt{1-\left(\frac{L}{L_c}\right)^2}\right).
\label{periapo}
\end{equation}

Of course in the above equation the $-$ sign gives the pericenter and the $+$ sign gives the apocenter.
In conclusion, once we have, as reference, a circular orbit of radius $r_c$ we may compare it with a set of orbits at same energy just varying the ratio $L/L_c$. The eccentricity of the orbit is, trivially, 

\begin{equation}
e=\frac{r_+-r_-}{r_-+r_+} = \sqrt{1-\left(\frac{L}{L_c}\right)^2}.
\label{ecc}
\end{equation}
 
We varied the parameter 

\begin{equation}
\alpha \equiv {\left(\frac{L}{L_c}\right)}^2,
\end{equation}

in order to sample GC orbits of different eccentricity and same orbital energy. Of course, $\alpha =0$ for radial orbits ($e=1$) and $\alpha = 1$ for circular orbits ($e=0$).

The cartesian reference frame has been chosen as that with the $x$-axis along the line connecting the GC with the SMBH and $y$-axis orthogonal, so that the $(x,y)$ frame is equiverse to the GC orbital revolution. 
 
In the restricted three-body problem, it is well known the existence of the Hill's surfaces which enclose the two finite-mass bodies \citep{sze66}. The radius of the Hill's sphere, given by
\begin{equation}
r_L=r_0{\left(\frac{M_{GC}}{3M_{BH}}\right)}^{1/3},
\end{equation}
defines the spherical volume around the GC where its gravitational potential dominates. Outside the Hill's sphere, the BH potential overcomes the one of the cluster. For the set of parameters used in our scattering experiments, $0.55$ pc $\le r_L \le 0.63$ pc, $1.19$ pc $\le r_L \le 1.36$ pc and $2.55$ pc $\le r_L \le 2.93$ pc for a GC of mass $10^4$ M$_{\odot}$, $10^5$ M$_{\odot}$ and $10^6$ M$_{\odot}$, respectively.

A meaningful study refers to the fate of stars moving around the GC with orbits initially all within the GC influence radius. Therefore, we put the initial circular orbits, on which the test star moves around the GC, inside this sphere by setting the radius of this orbit to be a fraction ($<1$) of the distance from the first Lagrange point (L1) and the GC.

For the sake of statistical significance, once fixed the unperturbed star circular orbit we sampled cases with initial different phases, in the range $0\div 360^\circ$ at increments of $15^\circ$ \citep*{gin12}.

To summarize, in this paper the values of the relevant initial parameters have been set as follows (see also Table 1): 

\begin{itemize}
\item the super massive black hole mass is $M_{BH}=10^8$ M$_{\odot}$;
\item the globular cluster mass, $M_{GC}$, assumes the three values $10^4$, $10^5$, and  $10^6$ M$_{\odot}$;
\item the test star mass, $m_{*}$, is set equal to $1$ M$_{\odot}$;
\item the GC reference circular orbit has the radius $r_0=10$ pc; 
\item the GC orbital eccentricity ranges from $e=0.71$ ($\alpha = 0.5$) to $e=0.95$ ($\alpha = 0.1$) and is parametrized varying $0.1\leq \alpha \leq 0.5$ at steps of $0.1$;
\item the test star circular orbit radius around the GC is parametrized by $\beta \equiv r/r_L$, whose values are in the range $0.08 \div 0.25$;
\item the star initial position on the circular orbit of given radius (see above) is parametrized by adopting a set of 24 different angles spanning $0\div 360^\circ$ with a $15^\circ$ step;
\item the star circular orbit around the GC and the GC orbit respect to the BH are coplanar.
\end{itemize}

The choice of the range of $\alpha$, and consequently of $e$, toward large values of $e$, is due to that (as we will see in the Results Section) the efficiency of the energy transfer on the test star orbiting the GC tends to vanish at eccentricities less than $\sim 0.6$.

\begin{table}
\caption{The values of $\alpha$, the eccentricity $e$, the pericentre $r_-$ and apocentre $r_+$ of the GC elliptical orbits.}
\centering
\begin{tabular}{c|c|c|c|}
\hline
$\alpha$ & $e$ & $r_-(pc)$ & $r_+ (pc)$ \\
\hline
0.1 & 0.95 & 0.51 & 19.6 \\
\hline
0.2 & 0.89 & 1.06 & 18.9 \\
\hline
0.3 & 0.84 & 1.63 & 18.5 \\
\hline
0.4 & 0.77 & 2.25 & 17.8 \\
\hline
0.5 & 0.71 & 2.93 & 17.1 \\
\hline
\end{tabular}
\label{tab1}
\end{table}

Given the above set of initial parameters, we integrated the system of the differential equations of the 3-bodies (SMBH, GC and star) motion

\begin{equation} \label{eqn:mot}
{\ddot{\textbf{r}}}_i=-G\sum\limits_{j\ne i}\frac{m_j(\textbf{r}_i-\textbf{r}_j)}{\left|\textbf{r}_i-\textbf{r}_j\right|^3},
\end{equation}

for $i=1, 2, 3$, using the fully regularized algorithm of \citet{mik01}. The need of a regularized algorithm is due to the enormous range of variation of the masses involved, going from $1$ M$_\odot$ of the test star to $10^8$ M$_\odot$ of the SMBH. Any not-regularized direct summation code would fail in dealing with the close star-SMBH interaction and would carry to an enormous energy error during the close triple encounter.

In the Mikkola's ARW code this problem is overcome by a transformed leapfrog method, which leads to extremely accurate integrations of the bodies trajectories when combined with the Bulirsch-Stoer extrapolation method \citep{bul66,mik99a,mik99b,mik06,mik08,hel10}. Thanks to the regularized algorithm, the fractional energy error is kept  below $10^{-11}$ over the whole integration time.

\section{Results}
\label{sec:res}

\begin{figure*}
\centering
\subfigure{\includegraphics[scale=0.8]{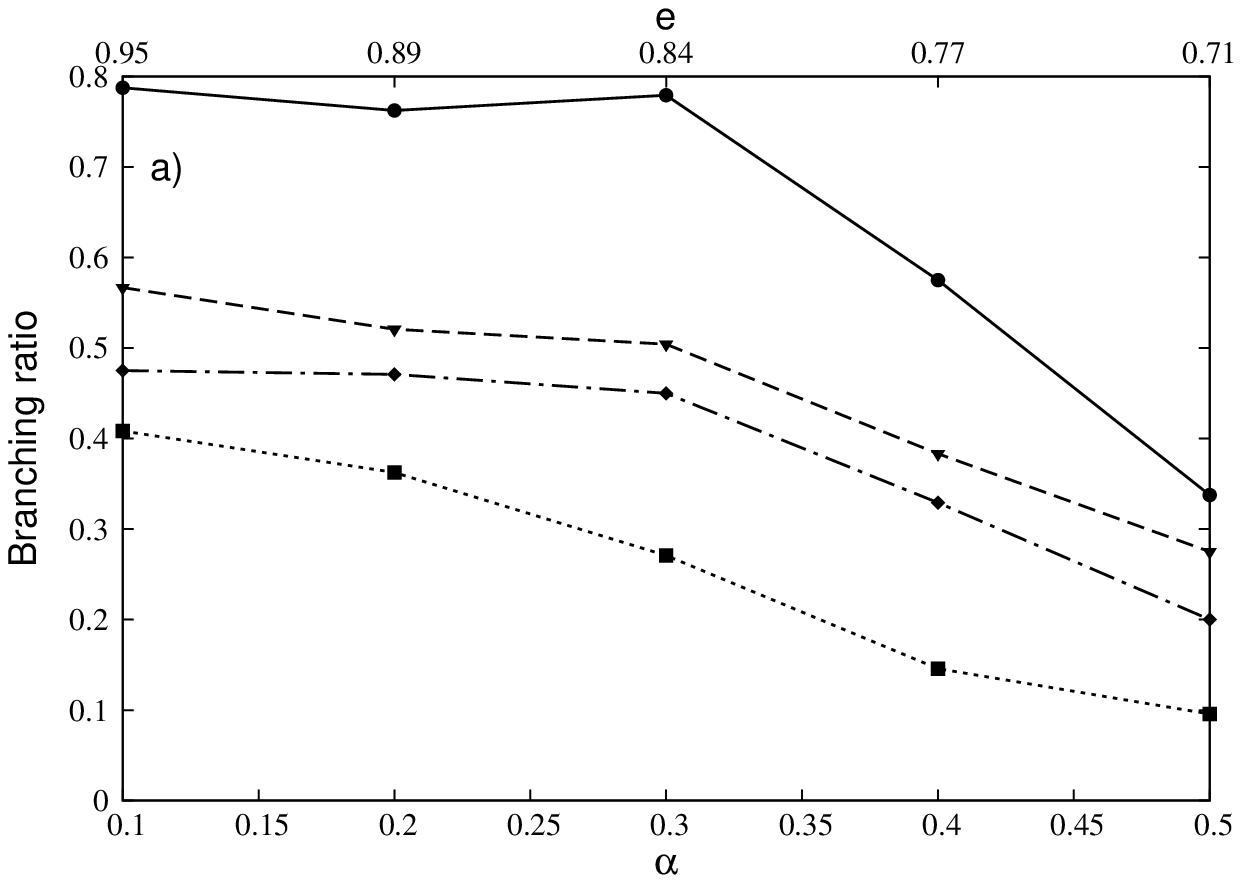}\label{fig:all_s}}
\subfigure{\includegraphics[scale=0.8]{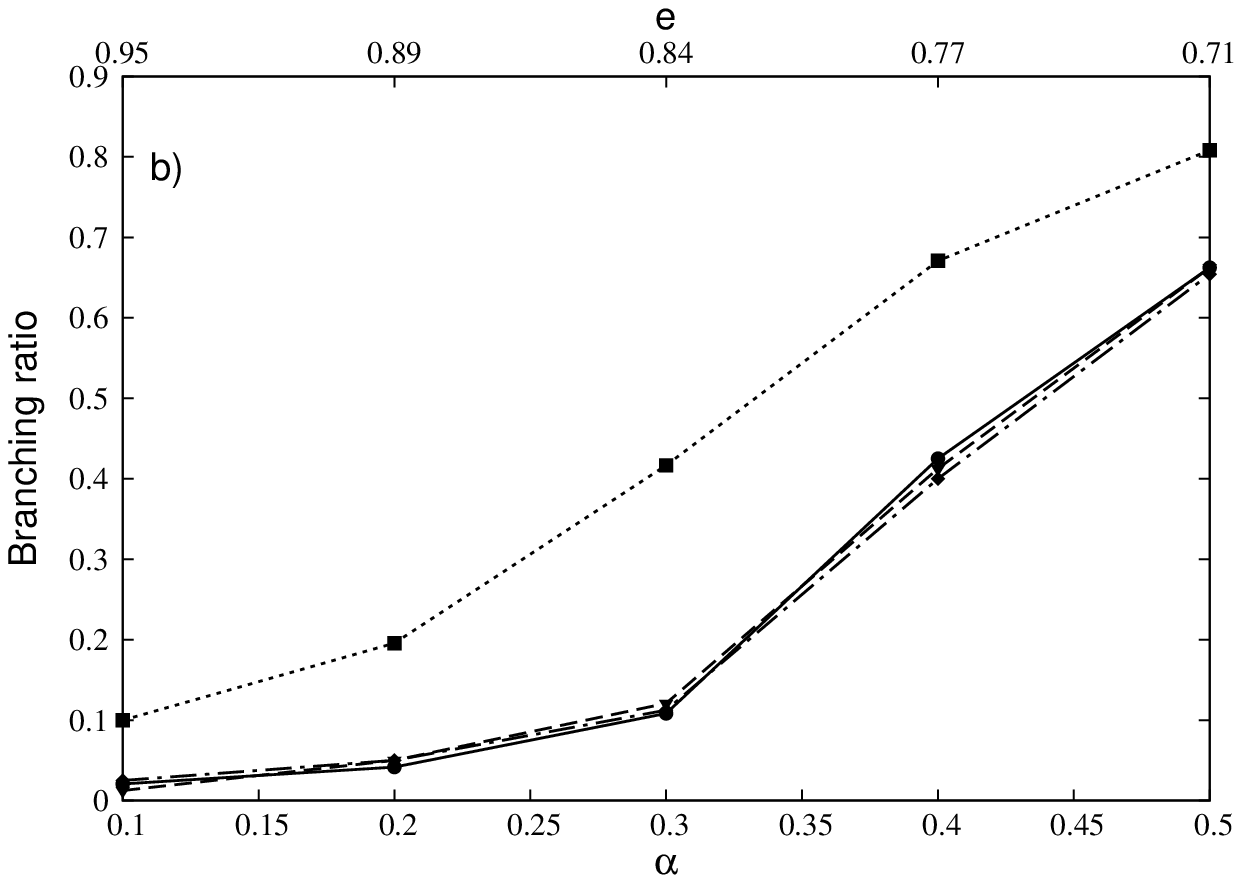}\label{fig:all_gc}}
\subfigure{\includegraphics[scale=0.8]{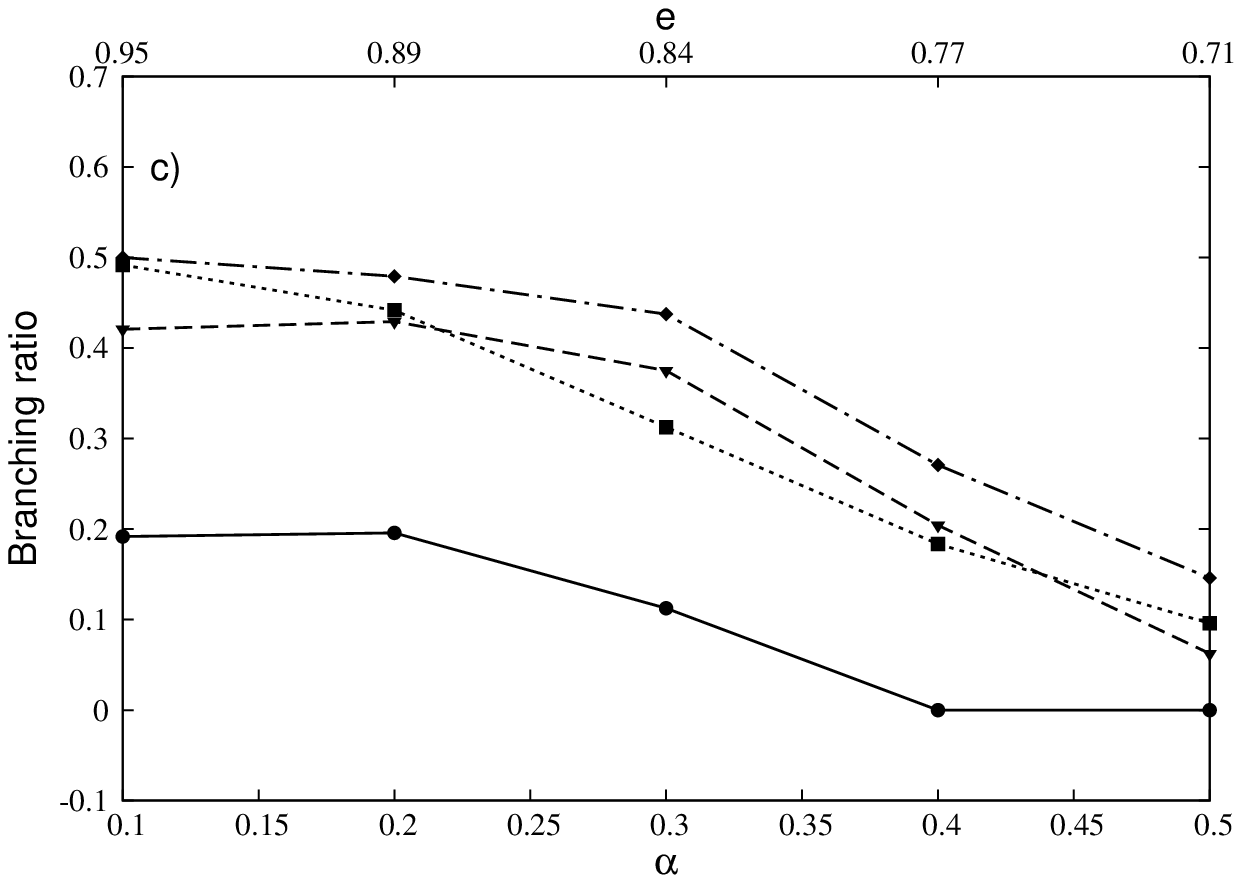}\label{fig:all_e}}
\caption{Branching ratios of stars captured by the BH (a), GC stars (b) and ejected stars (c), after GC-BH scattering, for $M_{GC}=10^4$ M$_{\odot}$ (solid line), $M_{GC}=10^5$ M$_{\odot}$ (dashed line), $M_{GC}=10^6$ M$_{\odot}$ (dot-dashed line) and different GC orbits, parametrized by $\alpha=(L/L_c)^2$. The dotted lines represent the branching ratios when the star orbit and GC orbit are perpendicular for $M_{GC}=10^6$ M$_{\odot}$.} 
\end{figure*}

\begin{figure*}
\centering
\begin{minipage}{20.5cm}
\subfigure{\includegraphics[scale=0.70]{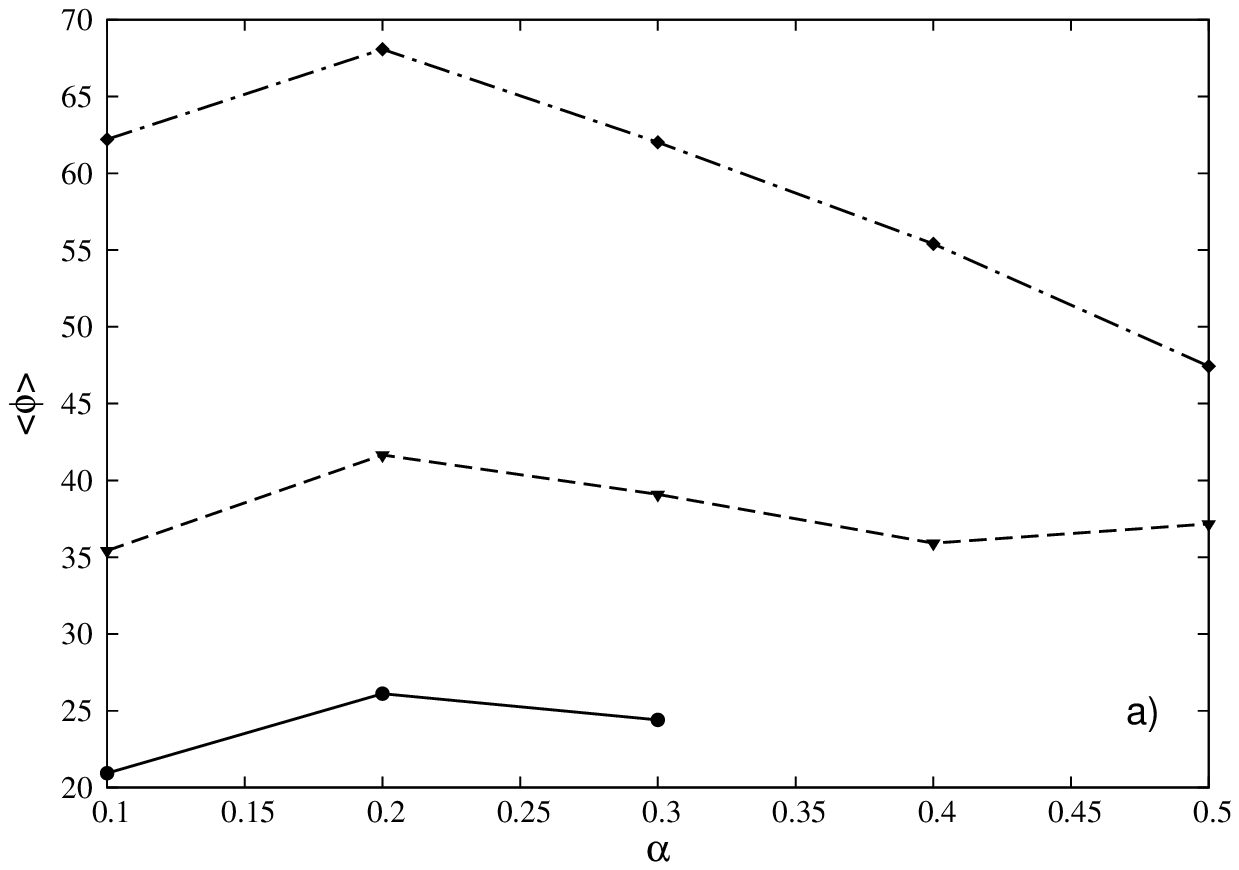}\label{fig:angle}}
\subfigure{\includegraphics[scale=0.70]{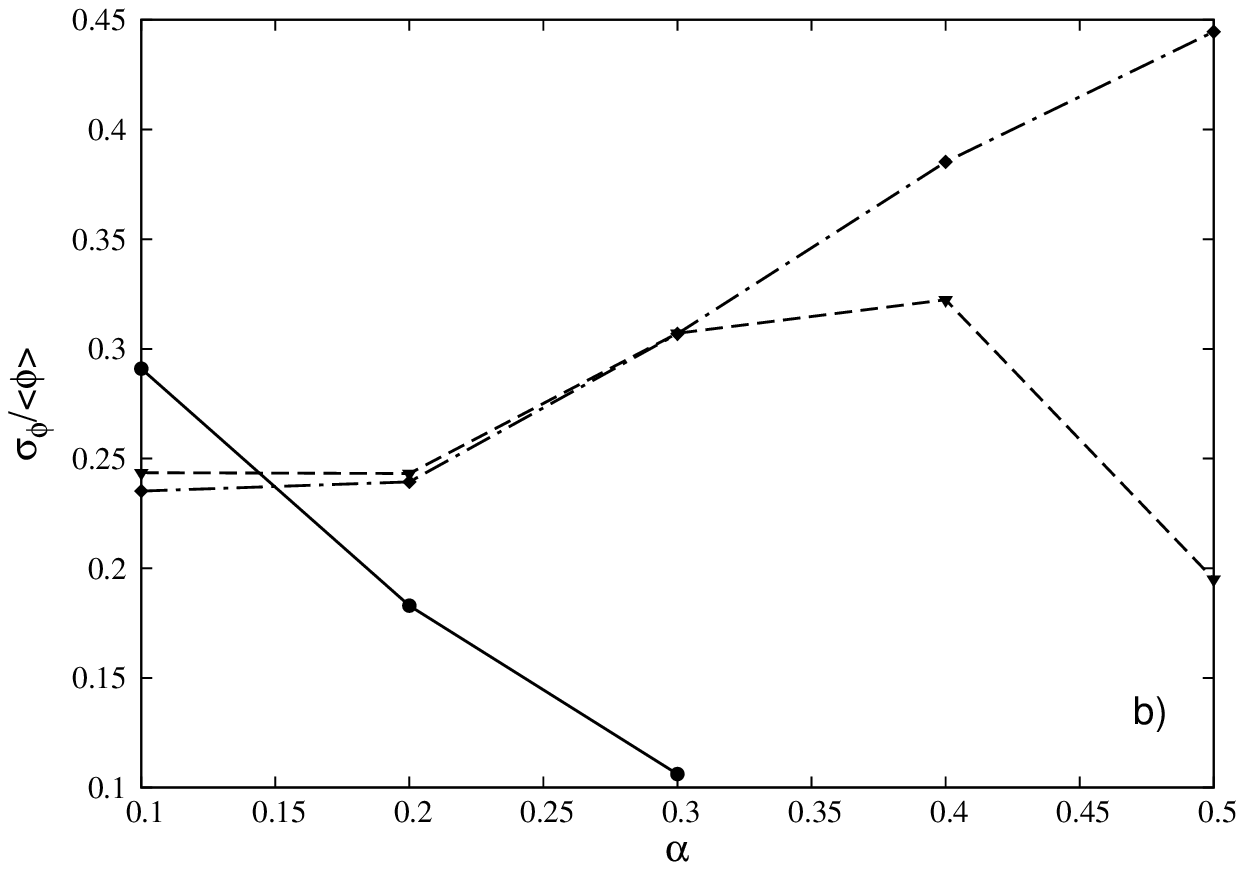}\label{fig:ang_err}}
\end{minipage}
\caption{Average ~\lq emission\rq~ angle of the ejected stars (a) and relative dispersion (b), both as function of $\alpha$, for $M_{GC}=10^4$ M$_{\odot}$ (solid line), $M_{GC}=10^5$ M$_{\odot}$ (dashed line), $M_{GC}=10^6$ M$_{\odot}$ (dot-dashed line).}
\end{figure*}

In our scattering experiments the test star orbiting the GC has three possible fates: 

\begin{enumerate}
\renewcommand{\theenumi}{(\arabic{enumi})}
\item it remains bound to the GC on an orbit significantly perturbed respect to the original one;
\item it becomes a {\it high velocity} star, either bound or unbound to the host galaxy;
\item it is captured by the massive BH gravitational field and starts revolving around it.
\end{enumerate}

The distinction among these three different situations is made by computing the mechanical energy of the star respect to the BH and the GC after the scattering. If the energy of the star respect to the GC remains negative, the star remains bound to the GC, while if this energy becomes positive, while the star energy respect to the Black Hole is negative, the star becomes bound to the BH. Finally, if both these energies are positive, the star is able to leave the BH-GC system and, according to the assumed galactic potential model, it will be bound or unbound respect to the galaxy. The branching ratios, i.e. the probability of different outcomes, are plotted in figs. \ref{fig:all_s},\ref{fig:all_gc},\ref{fig:all_e}.

Fig. \ref{fig:all_s} gives the branching ratio of stars which are captured by the BH after the GC-BH scattering and become bound to the BH, as function of $\alpha$ for the different GC masses. As expected, this ratio decreases for larger values of $\alpha$ (less eccentric orbits), as well as for larger GC masses.

Fig. \ref{fig:all_gc} shows the branching ratio of stars which remain gravitationally bound to the Globular Cluster after GC-BH scattering, as function of $\alpha$ for the different GC masses. The branching ratio increases for higher values of $\alpha$, i.e. for less eccentric orbits, and is almost independent of the GC mass in the range studied. Note that at values of $e\sim 0.7$ the fraction of bound stars is about 80\% for the $10^4$ M$_\odot$ GC and about 67\% for the $10^5$ and $10^6$ M$_\odot$ cases.

Finally, Fig. \ref{fig:all_e} shows the branching ratio of stars which, after GC-BH scattering, leave the BH-GC system becoming high velocity stars, as function of $\alpha$ for the different GC masses. This fraction decreases for larger values of $\alpha$ (less eccentric orbits) and increases for larger GC masses.

The Lagrangian radius have an important role during the close interaction between GC and BH. A star is lost by the GC when it crosses the Hill's surface through or near the first or the second Lagrangian point. If it passes through L1, its fate is the capture by the BH, while in the second case it will escape the whole system, becoming unbound both respect to the BH and the GC. The first channel is favoured by lower GC to BH mass ratios, since the BH potential is stronger and is able to capture a higher number of GC stars making them pass through the first Lagrangian point. At the same time, the GC gravitational potential is not so intense to give the star a velocity high enough to escape the whole GC-BH system. Therefore, the branching ratio for the production of stars captured by the BH is higher for lower GC masses, while the branching ratio of ejected stars increases for higher GC masses.

Another important feature of this mechanism is the significant level of collimation of the ejected stars. Figure \ref{fig:angle} shows the average ejection angle $\phi$ as function of $\alpha$ for different GC masses. This angle is that between the velocity vector at ejection and the $x$ axis of the inertial reference frame, taken in the direction pointing from the BH to the initial position of the GC. Figure \ref{fig:ang_err} shows the ratio between the standard deviations $\sigma_{\phi}$ and $\phi$, giving a measure of the ejection collimation. If the value of $\sigma_{\phi}$ was precisely zero, it would mean that all the stars are ejected along the same direction after the GC-BH interaction. Therefore, non-zero values measure the width of the loss-cone within which stars are ejected. We see that the average emission angle increases for larger values of the GC mass, while the level of collimation decreases. Therefore, stars ejected during the interaction of a low-mass GC and a massive BH are concentrated in a small amplitude jet, which, instead, has a greater span for a high mass GC. This results are compatible with \citet{ses06}, who found, studying the HVS production of a binary BH, that the collimation is higher when the mass ratio between the BHs is lower. Actually, we found that stars are highly collimated in jets when $M_{GC}=10^4$ M$_{\odot}$, while for higher GC masses the ejection jets tends to become a nearly isotropic emission.

\subsection{High velocity stars}
\label{hvs}
\begin{figure*} 
\centering
\begin{minipage}{20.5cm}
\subfigure{\includegraphics[scale=0.72]{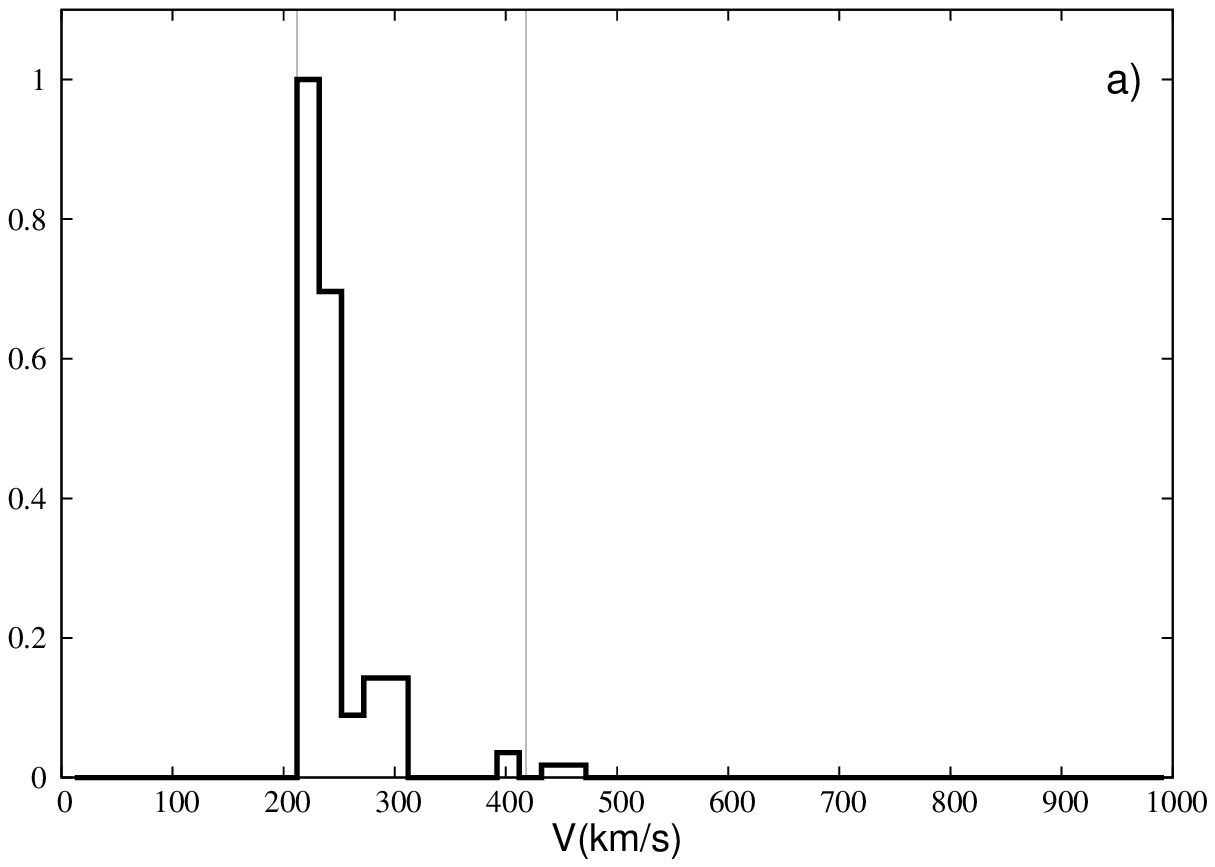}\label{fig:ist_vel_4el}}
\subfigure{\includegraphics[scale=0.72]{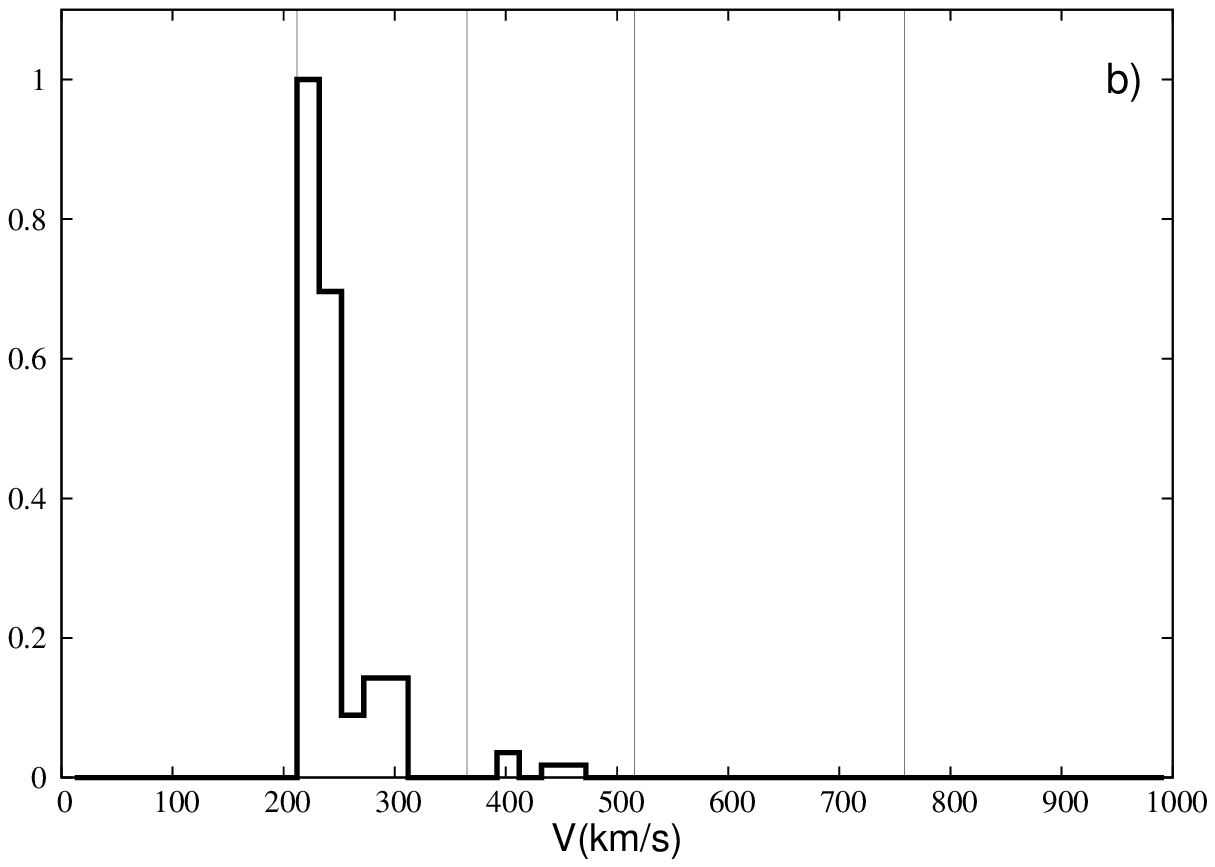}\label{fig:ist_vel_4sp}}
\end{minipage}
\begin{minipage}{20.5cm}
\subfigure{\includegraphics[scale=0.72]{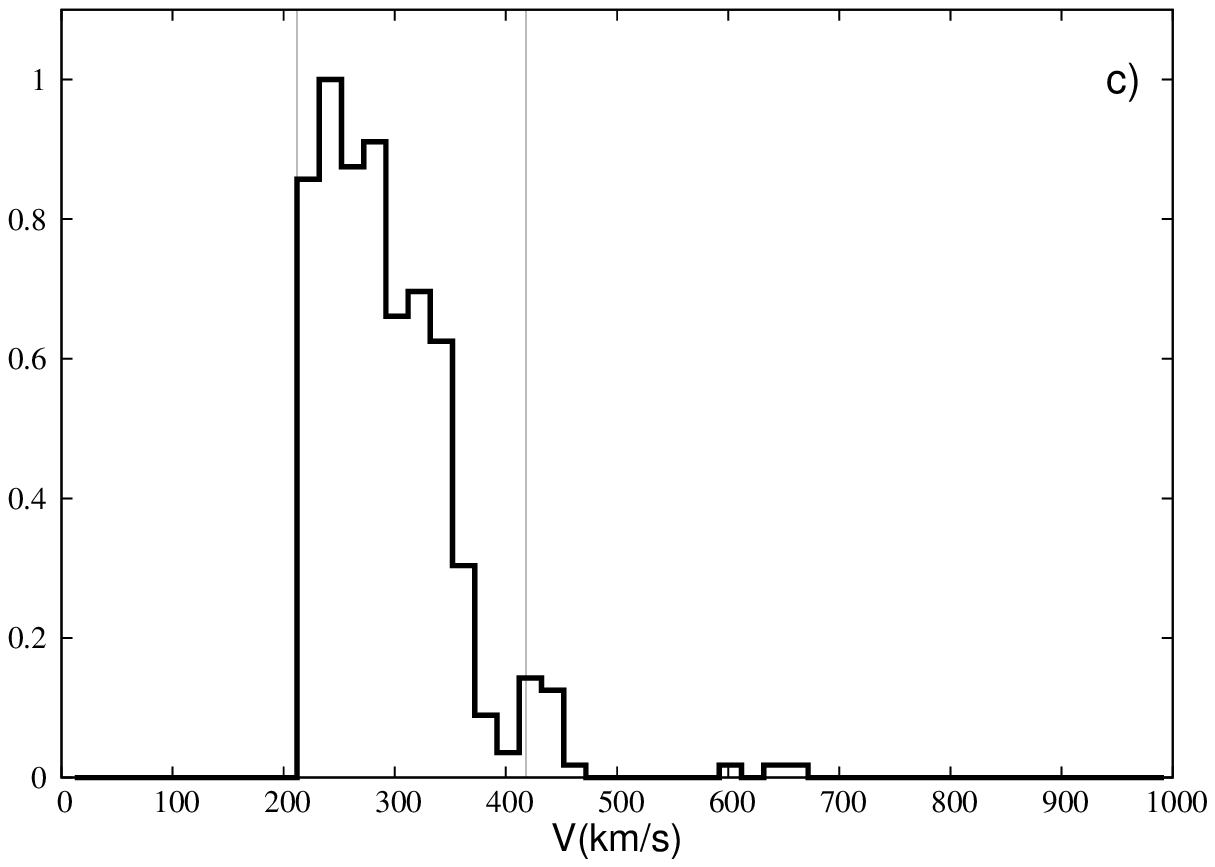}\label{fig:ist_vel_5el}}
\subfigure{\includegraphics[scale=0.72]{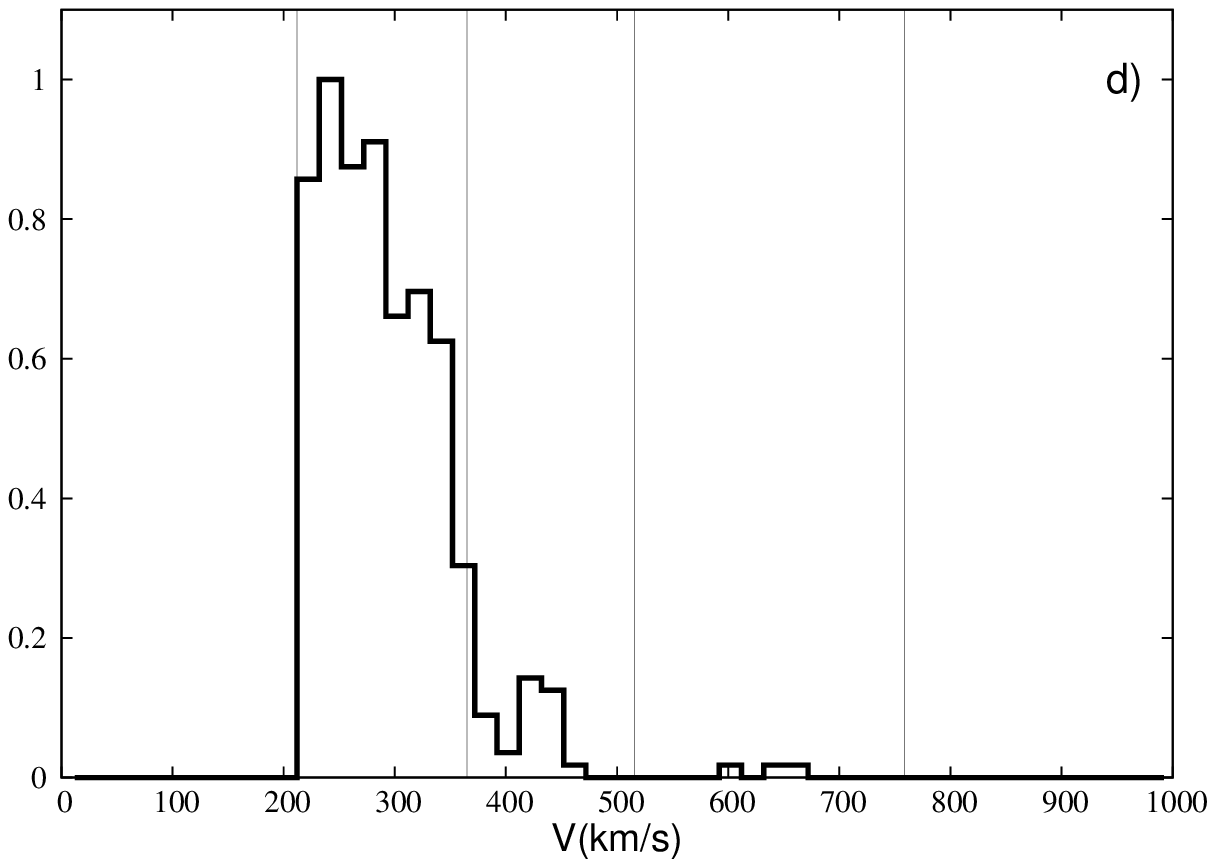}\label{fig:ist_vel_5sp}}
\end{minipage}
\begin{minipage}{20.5cm}
\subfigure{\includegraphics[scale=0.72]{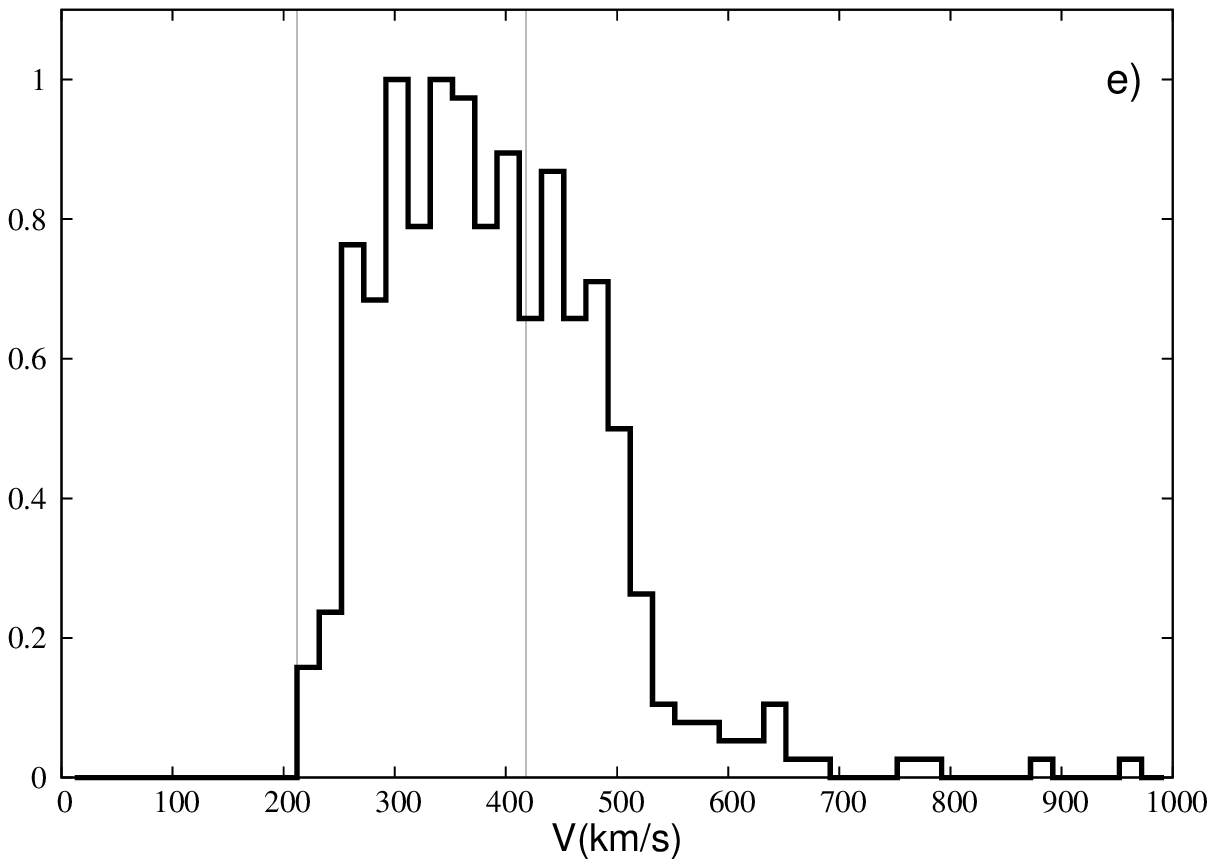}\label{fig:ist_vel_6el}}
\subfigure{\includegraphics[scale=0.72]{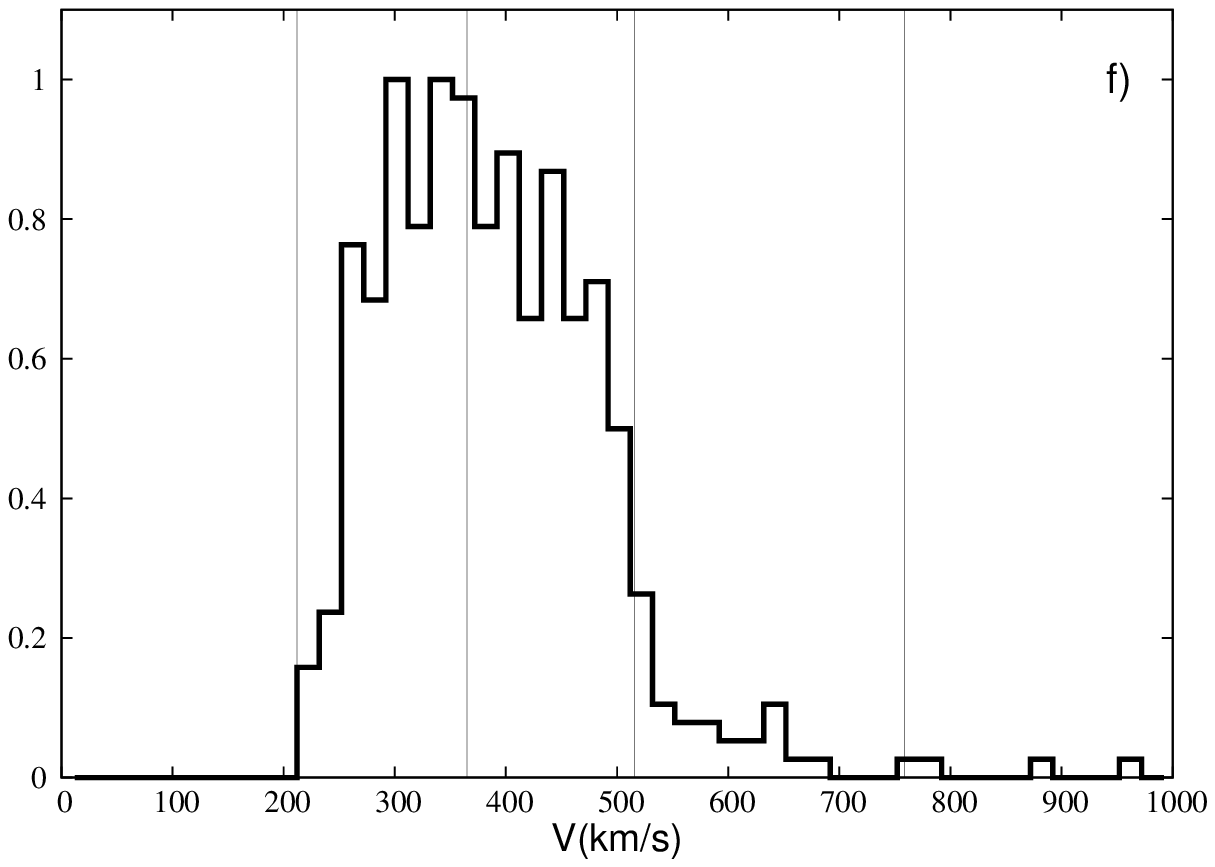}\label{fig:ist_vel_6sp}}
\end{minipage}
\caption{Velocity distribution of escaping stars for $M_{GC}=10^4$ M$_\odot$ (a,b), $M_{GC}=10^5$ M$_\odot$ (c,d) and $M_{GC}=10^6$ M$_\odot$ (e,f) and all the orbits, both for a M$_{tot}=7.81 \times 10^{10}$ M$_{\odot}$ elliptical galaxy \citep{mar03} (left column) and a M$_{tot}=6.60 \times 10^{11}$ M$_{\odot}$ spiral galaxy \citep{fuj09} (right column). Vertical lines indicate the escape velocity of the various galactic components (see Sect. \ref{hvs})}
\label{fig:distr}
\end{figure*}

From our scattering experiments, it is possible to derive the velocity profile of the ejected stars. Fig. \ref{fig:distr} shows the velocity profile of the ejected stars at $20$ pc (as it will be justified in the following) for different GC masses and all the orbits. The distributions are nearly Gaussians, cut at $212$ km s$^{-1}$, which is the escape velocity from the BH. Moreover, the peak of the distribution depends on the GC mass, the greater the velocity peak the larger the GC mass. The ejected stars can have two fates:

\begin{enumerate}
\renewcommand{\theenumi}{(\arabic{enumi})}
\item to remain gravitationally bound to the galaxy, although having escaped the BH-GC system;
\item to become unbound stars, and so HVS, if their kinetic energy is sufficient to overcome the galaxy gravitational potential well.
\end{enumerate}

To evaluate whether stars formerly belonging to the GC and ejected at high velocity remain bound to the host galaxy, an assumption on the galactic field has to be made. In our analysis, we assumed two different models for the host Galaxy, one as an elliptical and one as a spiral galaxy.

The elliptical galaxy potential is represented as a two-component model given by a spherical bulge-halo summed to the SMBH potential. The bulge-halo potential is given by \citep{her90}

\begin{equation}
\Phi_{b}(r)=-\frac{GM_{b}}{r+a_b},
\end{equation}

where $M_{b}=7.8\times 10^{10}$ M$_{\odot}$ and $a_b=5.4$ kpc. Note that these values are taken from \citet{mar03} to represent  the elliptical galaxy NGC 3377, whose central BH has an estimated mass $M_{BH}={1.0}^{+0.9}_{-0.1}\times 10^8$ M$_{\odot}$.

Assuming the host galaxy as a spiral, we consider a four-component model for its  potential

\begin{equation} 
\Phi(r,z) =\Phi_{BH}(r)+\Phi_{b}(r)+\Phi_{d}(r,z)+\Phi_{h}(r),
\end{equation}

where the indexes $b,d,h$ stands for bulge, disk and halo, respectively. As above, the  bulge potential is expressed as a Hernquist sphere \citep{her90} with the constants $M_{b}=10^{10}$  M$_{\odot}$ and $a_B=1$ kpc, as taken from \citet{krn08} and \citet{lin14}, to reproduce the parameters observed in giant disk galaxies such as those described by \citet*{wan92} and \citet*{rwn94}.

The axisymmetric disk potential is \citep{miy75}
 
\begin{equation}
\Phi_{d}(r,z)=-\frac{GM_{d}}{\sqrt{r^2+\left(a_d+\sqrt{z^2+b_d^2}\right)^2}},
\end{equation}

where $M_{d}=10^{11}$ M$_{\odot}$, $a_d=6.5$ kpc and $b_d=0.26$ kpc \citep{lin14}.

Finally, the halo density distribution accounts for the presence of a spherical dark matter halo, of total mass $5.6\times 10^{11}$ M$_{\odot}$, whose potential is \citep{bin81}

\begin{equation}
\Phi_{h}(r)=\frac{1}{2}v_h^2\ln(r^2+r_h^2) + c,
\end{equation}

where $v_h=250$ kms$^{-1}$, $r_h=2$ kpc and $c$ is a constant which gives a match with an external ($r \geq 50$ kpc) keplerian potential \citep{fuj09,lin14}.

Given a galaxy model, it is possible to calculate the escape velocity
\begin{equation} \label{eqn:vesc}
v_{esc}^{(j)}(r,z)=\sqrt{\sum_i -2\Phi_i(r,z)},
\end{equation}

which depends, besides on the galaxy model itself ($j$=E,S), also on the position in which it is computed. In our 3-body scattering, we follow the trajectories of stars until they are $20$ pc far from the BH and compute their velocity and the escape velocity at this distance. At this distance the contribution of the galaxy components is not negligible in the Eq. \ref{eqn:vesc}, but its gravitational potential is nearly constant, both for the elliptical and spiral galaxy. Therefore, the motion of the escaping star is due to only the central BH.

If $v_*<v_{esc}$, the star, although escapes the BH-GC system, will be gravitationally bound to the galaxy. In this case, the star can be bound only to some galactic components, as specified by the vertical lines in Fig. \ref{fig:distr}, which divide the distribution in different portions. The leftmost thick line in all the panels indicates the escape velocity from the BH ($212$ km s$^{-1}$). In the left column panels, the other vertical line refers to the escape velocity ($418$ km s$^{-1}$) from the BH +  bulge-halo system, while, in the right column panels, the other vertical lines indicate the escape velocity respect to the BH + bulge ($365$ km s$^{-1}$), BH + bulge + disk ($516$ km s$^{-1}$), BH + bulge + disk + dark halo ($759$ km s$^{-1}$), respectively. Therefore, for example, the portion of velocity distribution, on the right of the $365$ km s$^{-1}$ vertical line (for a spiral galaxy), is unbound respect to the BH + bulge component but bound respect to the disk and the dark halo.

On the contrary, if $v_*>v_{esc}$, it will become a HVS. The branching ratios of HVS, i.e. the probability of producing unbound stars from the galaxy with respect to the total ejected stars, are listed in Tab. \ref{tab:hvs} for all the orbits. The results, which depend on both the total mass of the host galaxy and on the shape of its gravitational potential, show that the higher is the GC mass the higher is the probability of producing HVS. Furthermore, according to the parameters chosen for the host galaxies gravitational potential, while the $10^6$ M$_\odot$ GC is able to generate HVS in both the galaxies, the $10^4$ M$_\odot$ and $10^5$ M$_\odot$ GC are able to produce HVS only in the elliptical galaxy.

\subsection{The role of star orbital inclination}

In our scattering experiments the initial orbit of the star and the orbit of the GC are coplanar. In order to check the effect of the relative inclination between the star and the GC orbit, we performed the same set of simulations for the $M_{GC}=10^6$ GC presented above in the case of star orbits initially lying on a plane perpendicular to the GC orbital plane. The resulting branching ratios are plotted in Fig. 1. While the branching ratios of the ejected stars and of the stars captured by the BH decrease, the branching ratio of stars which remain bound to the GC increases. Therefore, the overall effect is that stars tend to remain more bound to the cluster in the inclined case respect to the coplanar one.

\begin{figure}
\centering
\includegraphics[scale=0.7]{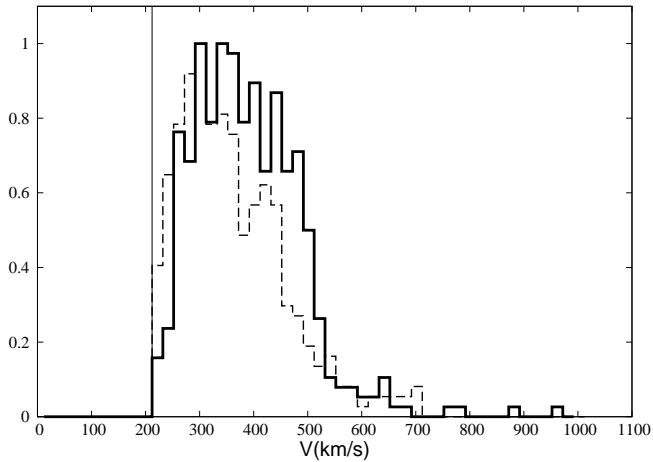}
\caption{Comparison between the velocity distributions of escaping stars for $M_{GC}=10^6$ M$_{\odot}$ when the GC orbit and star orbit are coplanar (solid line) and when they are perpendicular (dashed line). The distributions are cut on the left side at $212$ km s$^{-1}$, which corresponds to the escape velocity respect to the BH.}
\end{figure}

Fig. 4 shows the comparison of the velocity profiles. The perpendicularity of the orbits makes the distribution to peak at lower values of the velocity and the area under the distribution is smaller, because the branching ratio of ejected stars is lower than in the coplanar case.

\subsection{The role of a smooth GC potential}

In order to see the effect of a GC mass profile in the results, we performed the same set of simulations performed in the case of a $M_{GC}=10^6$ point mass GC, assuming a \citet{plu11} mass profile

\begin{equation}
M(r)=M_{GC}\frac{r^3}{{(r^2+a^2)}^{3/2}},
\end{equation}

where $M_{GC}$ is the total mass of the GC and $a$ its core radius, which is set to $0.5$ pc. Fig. 5 shows the velocity profiles for a point mass GC and a Plummer GC. The clear effect of smoothing the GC potential is that  the velocity distribution shifts towards lower values of the velocity. Actually, for the set of parameters chosen in this study the gravitational energy of the star is $\sim GM_{GC}/a$, which leads the peak of the nearly Gaussian distribution to a lower velocity and makes its dispersion decrease. Actually, if the GC is taken to be a point mass, for same radius of the circular orbit, the generic star of our simulation has a lower (more tightly bound) gravitational energy respect to the case of a GC smooth potential. Then the amount of gravitational energy that could be converted into kinetic energy would be higher, giving a larger number of ejected stars and a velocity distribution peaked at higher velocities. Clearly, this same effect of reduction of the efficiency in the star acceleration after the GC-MBH fly-by is obtained when the black hole mass is reduced, reducing thus the quantity of gravitational energy to inject in the test star motion. This means that we would expect a scaling of the phenomenon efficiency almost linear with the MBH mass, which would mean a reduction of the effects studied in this paper, where the SMBH mass is $10^8$ M$_\odot$, for a factor $0.04$ in the case of the Milky Way $4\times 10^6$ M$_\odot$ Sgr A* BH. This is just a rough qualitative sketch of a context, that of the scattering around the MW central MBH, that deserves a much more careful study which we will do in a forthcoming paper.
\begin{figure}
\centering
\includegraphics[scale=0.7]{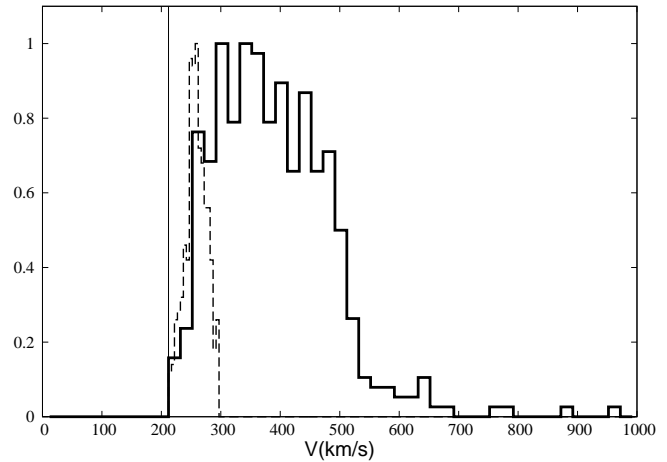}
\caption{Comparison between the velocity distributions of escaping stars for $M_{GC}=10^6$ M$_{\odot}$ when the GC is approximated as a point mass (solid line) and when it has a Plummer density profile with core radius $a=0.5$ pc (dashed line). The distributions are cut on the left side at $212$ km s$^{-1}$, which corresponds to the escape velocity respect to the BH.}
\end{figure}

\begin{table} 
\caption{Branching ratio (3rd column) of the unbound stars (HVS) respect to the total ejected stars for different galaxy models (E = elliptical, S = spiral).}
\centering
\begin{tabular}{c|c|c}
\hline
$M_{GC}$(M$_\odot$) & Galaxy & BR\\
\hline
$10^4$ & E & $1.67 \times 10^{-2}$\\
\hline
$10^4$ & S & $0$\\
\hline
$10^5$ & E & $4.47 \times 10^{-2}$\\
\hline
$10^5$ & S & $0$\\
\hline
$10^6$ & E & $0.35$\\
\hline
$10^6$ & S & $9.09 \times 10^{-3}$\\
\hline
\end{tabular}
\label{tab:hvs}
\end{table}

\subsection{Number of ejected stars}
\begin{figure} \label{fig:models}
\centering
\subfigure{\includegraphics[scale=0.65]{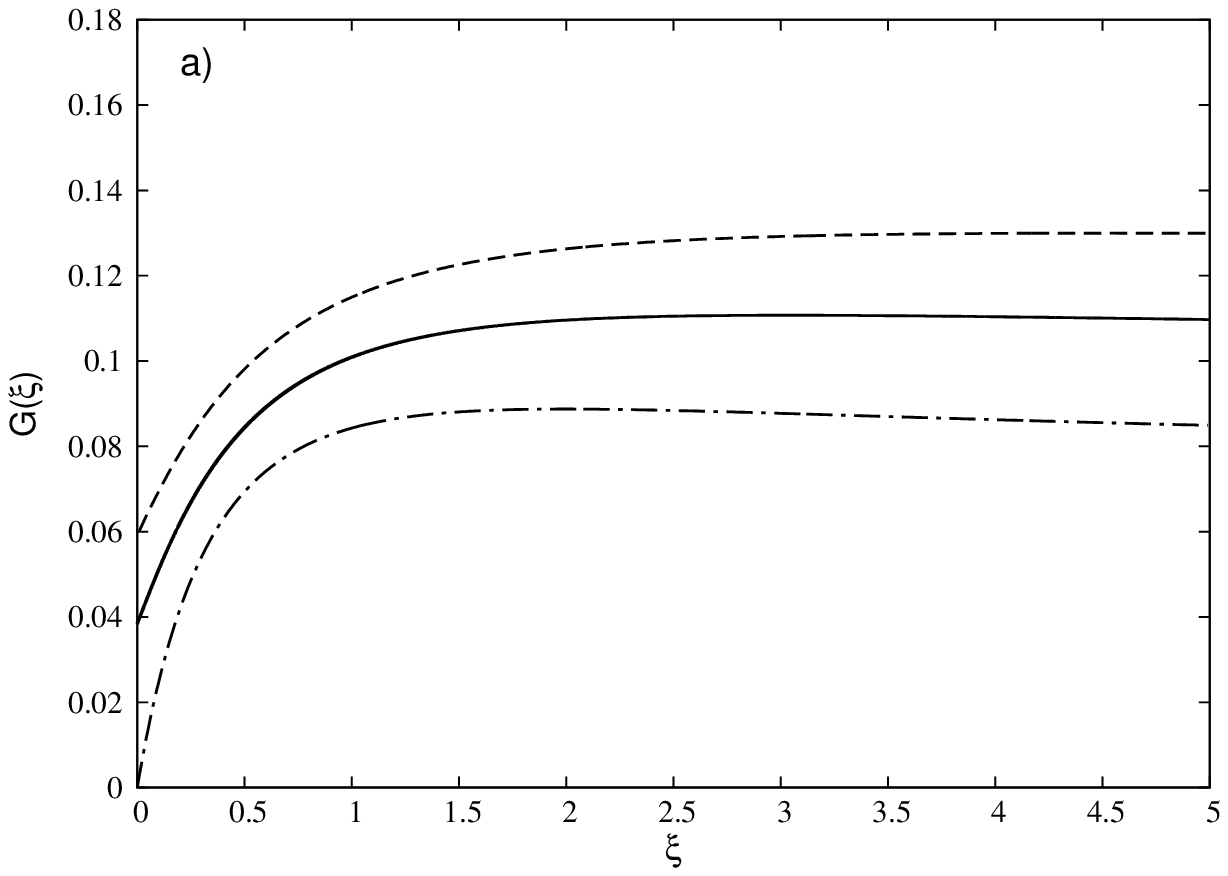}\label{fig:model_her}}
\subfigure{\includegraphics[scale=0.65]{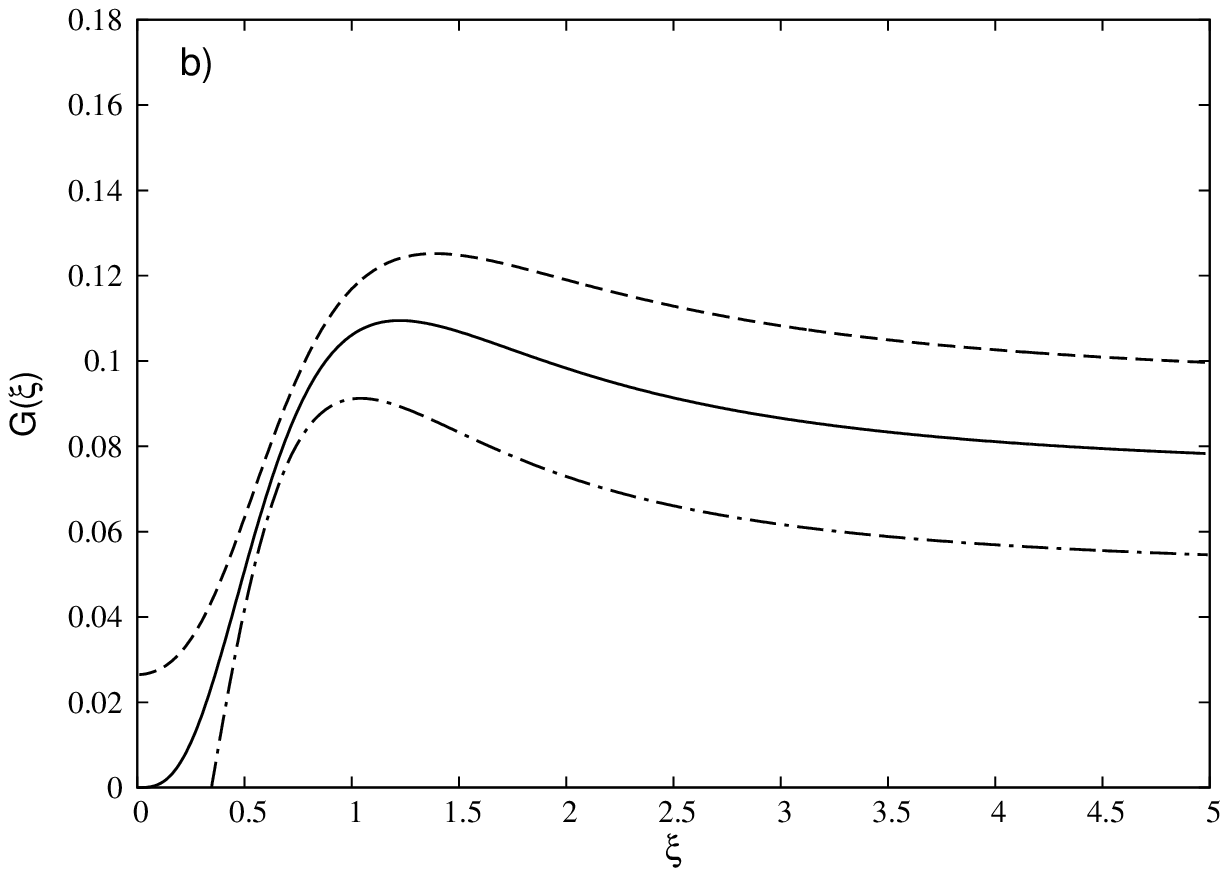}\label{fig:model_plu}}
\caption{Local fraction $G(\xi)\equiv \nu_{\mathrm{c}}(\xi)/\nu(\xi)$ of stars in nearly circular orbit with $\delta=0.05$, for the Hernquist (a) and the Plummer (b) potentials. The isotropic fraction ($\beta=0$, solid line) stands between the tangentially biased fraction ($\beta=- 1/2$, dashed line) and the radially biased fraction ($\beta=+ 1/2$, dot-dashed line).}
\end{figure}

After discussing the fate of an indivudal test star moving circularly around a point like GC, we want here to quantify the actual number of stars belonging to a GC that can become bound to the SMBH or ejected at high or even hyper velocity after the interaction with the SMBH.
A suitable estimation comes from the evaluation of the number of stars in nearly circular orbits in a self consistent model of GC of known distribution function (DF). Of course, the DF is not unique and depends both on the functional form of the GC gravitational potential and on the level of anisotropy. Here we consider two different gravitational potentials for the GC

\begin{itemize}
\item a Hernquist potential \citep{her90}

\begin{equation}
\Phi(r)=-\frac{GM}{r+a},
\end{equation}

where $M$ is the total GC mass and $a$ its core radius;

\item a Plummer potential \citep{plu11}

\begin{equation}
\Phi(r)=-\frac{GM}{\sqrt{r^2+a^2}},
\end{equation}

where, again, $M$ is the total GC mass and $a$ its core radius.

\end{itemize}

Subsequently, we treated both the isotropic and anisotropic cases. With the usual definition of the anisotropy parameter 

\begin{equation}
\beta=1-\frac{\sigma_{\theta}^2+\sigma_{\phi}^2}{2\sigma_r^2},
\end{equation}

where $\sigma_r$, $\sigma_{\theta}$, $\sigma_{\phi}$ are the velocity dispersions in spherical polar coordinates, an isotropic model has $\beta=0$, while $\beta$ is non-zero for  anisotropic cases. 

Using different modelisations, we computed the 
local fraction of stars of the GC on nearly circular orbits, $G(\xi)=\nu_c(\xi)/\nu(\xi)$, as function of $\xi\equiv r/a$, where 
$\nu(\xi)$ is the local density of bound stars of any velocity, while  $\nu_c(\xi)$ is that of stars moving on nearly circular orbit of scaled radius $\xi$, i.e. those with velocity very close to the local circular velocity $v_c(\xi)$. 

For the Hernquist potential, 

\begin{equation}
\nu(\xi)=\frac{M}{2\pi a^3}\frac{1}{\xi(1+\xi)^{3}},
\end{equation}

while for the Plummer potential

\begin{equation}
\nu(\xi)=\frac{3M}{4\pi a^3} \frac{1}{(1+\xi^2)^{5/2}}.
\end{equation}

The function $\nu_c(\xi)$ depends, besides the assumed potential, also on the degree of velocity anisotropy and can be  calculated as described in Appendix. The resulting $\nu_c(\xi)$ will depend also on the~\lq tolerance\rq~$\delta$, which quantifies the departure from the exact circular velocity. In our calculations we consider in $\nu_c(\xi)$ all the stars having a local speed in the interval $-\delta \leq v/v_c \leq +\delta$, with $\delta =0.05$.

As expected, Figs. \ref{fig:model_her}-\ref{fig:model_plu} show that the isotropic ($\beta = 0$) model fraction stays between the radially biased ($\beta=+1/2$) and the tangentially biased ($\beta=-1/2$) models.

\begin{figure*}
\centering
\begin{minipage}{20.5cm}
\subfigure{\includegraphics[scale=0.72]{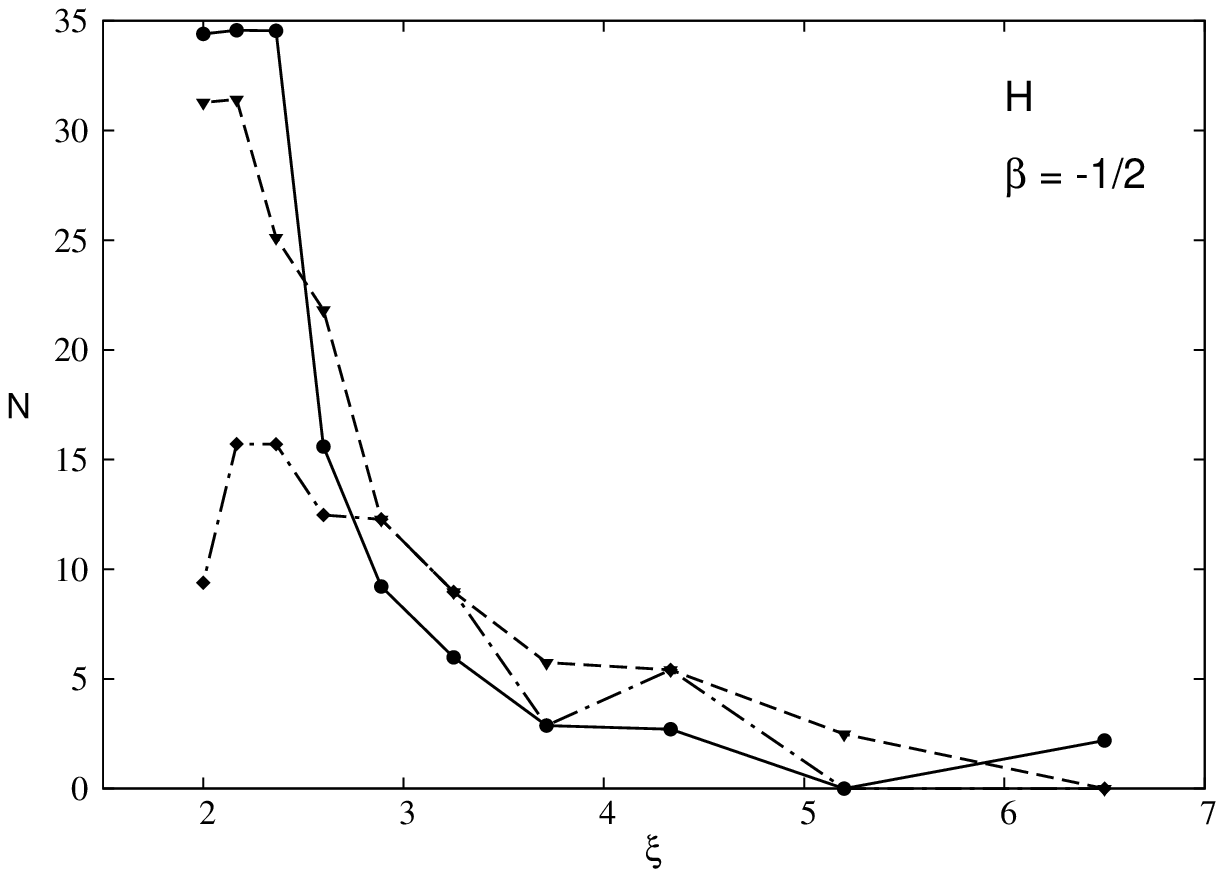}\label{fig:h_min_m4}}
\subfigure{\includegraphics[scale=0.72]{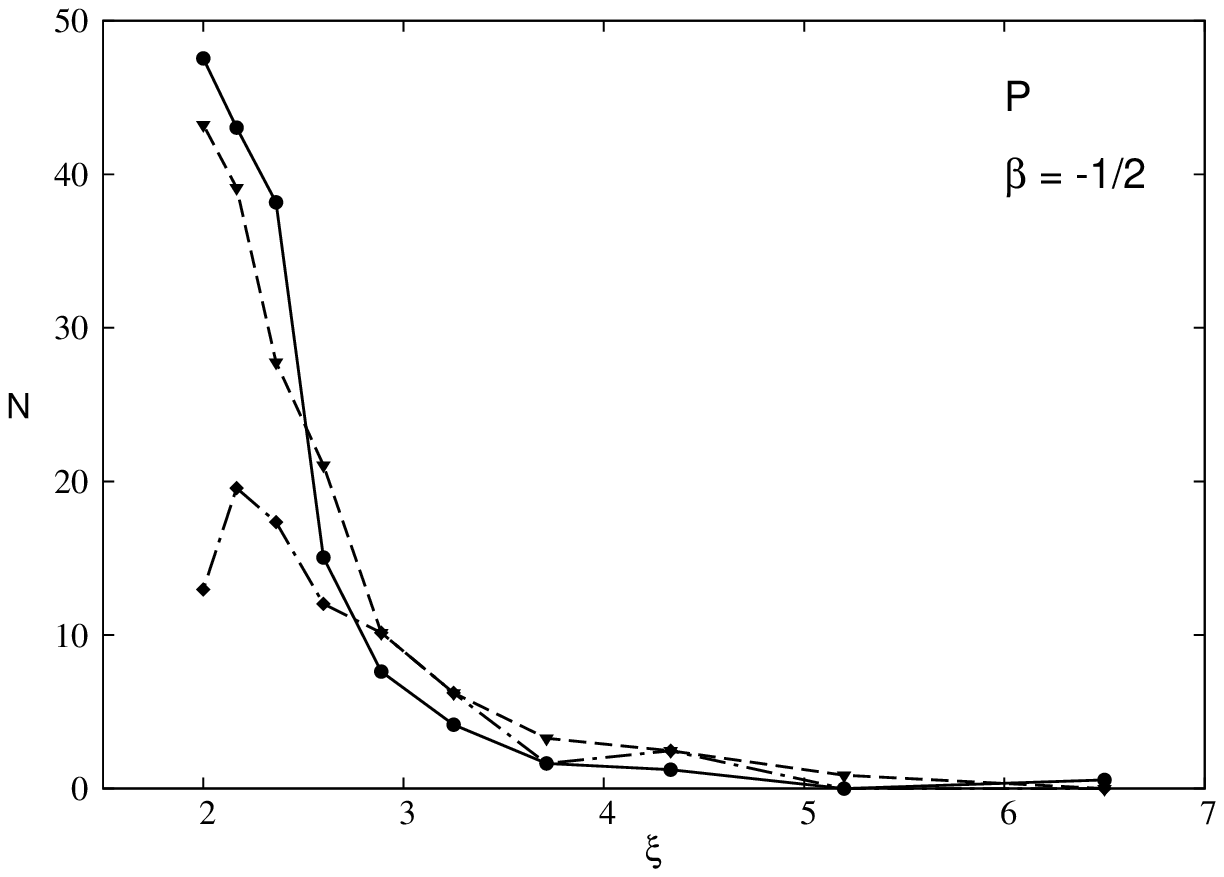}\label{fig:p_min_m4}}
\end{minipage}
\begin{minipage}{20.5cm}
\subfigure{\includegraphics[scale=0.72]{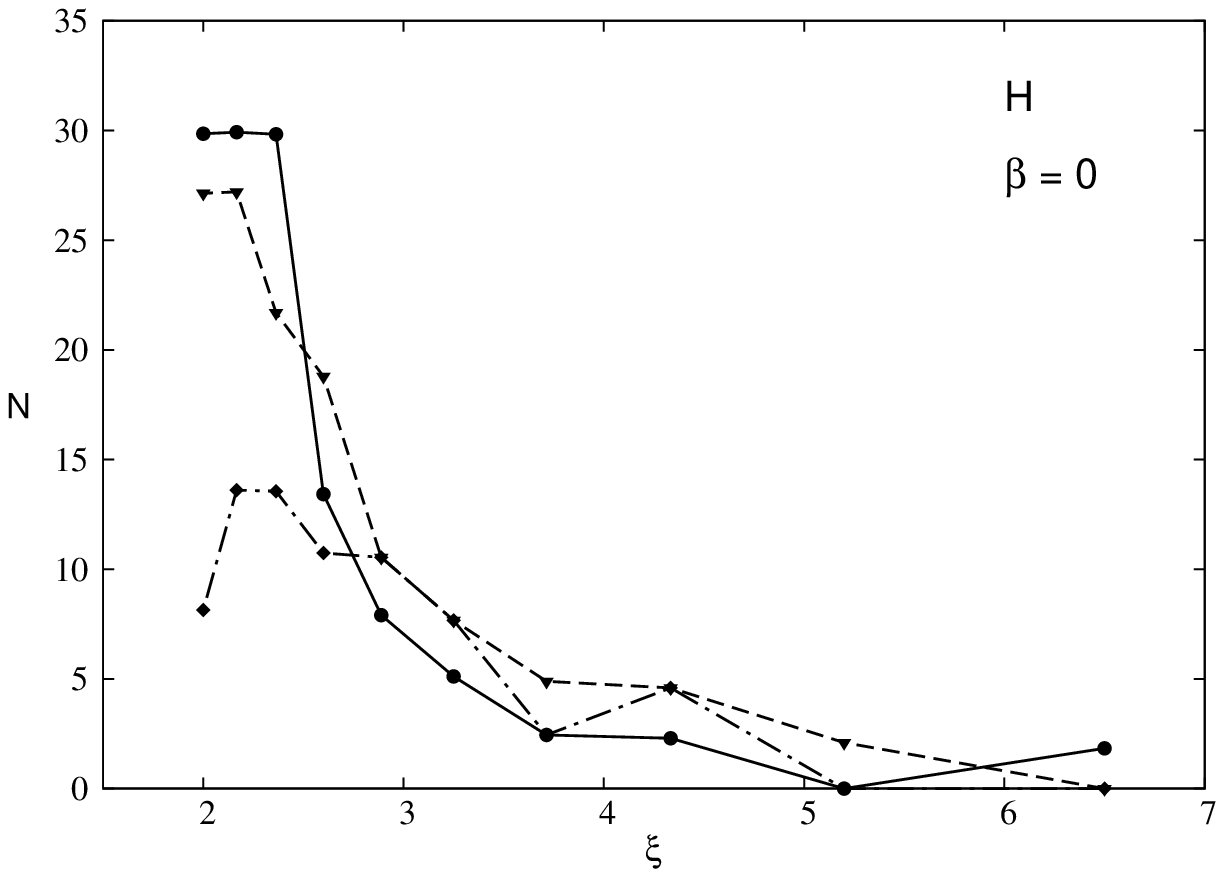}\label{fig:h_zer_m4}}
\subfigure{\includegraphics[scale=0.72]{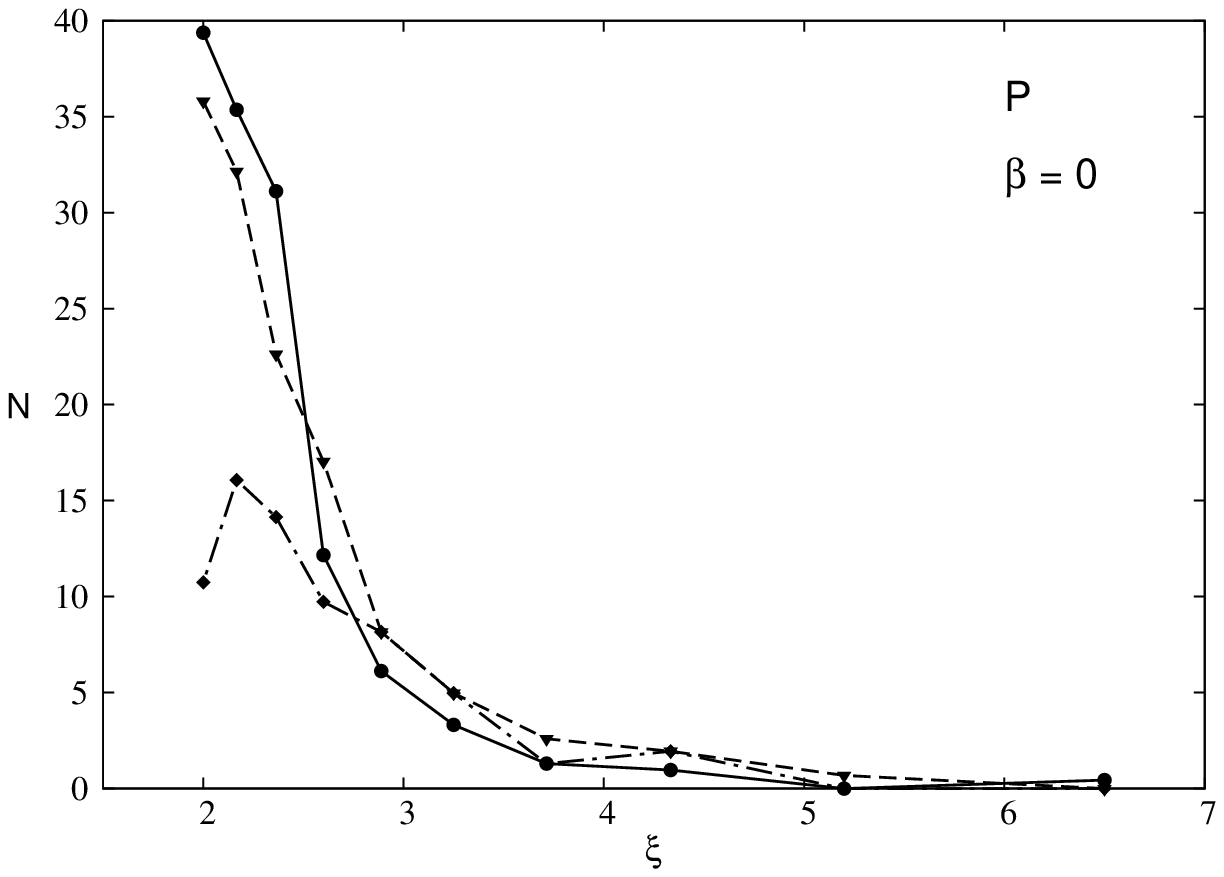}\label{fig:p_zer_m4}}
\end{minipage}
\begin{minipage}{20.5cm}
\subfigure{\includegraphics[scale=0.72]{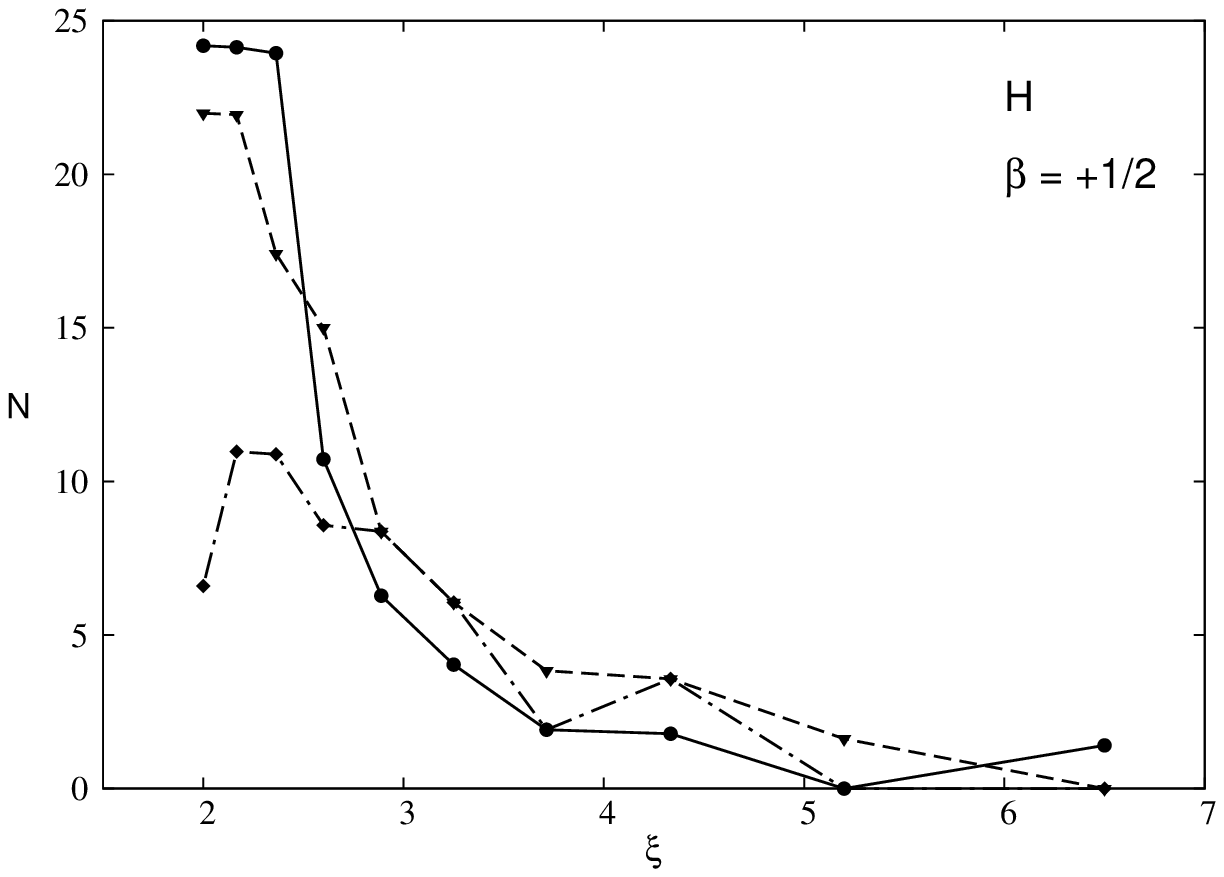}\label{fig:h_pl_m4}}
\subfigure{\includegraphics[scale=0.72]{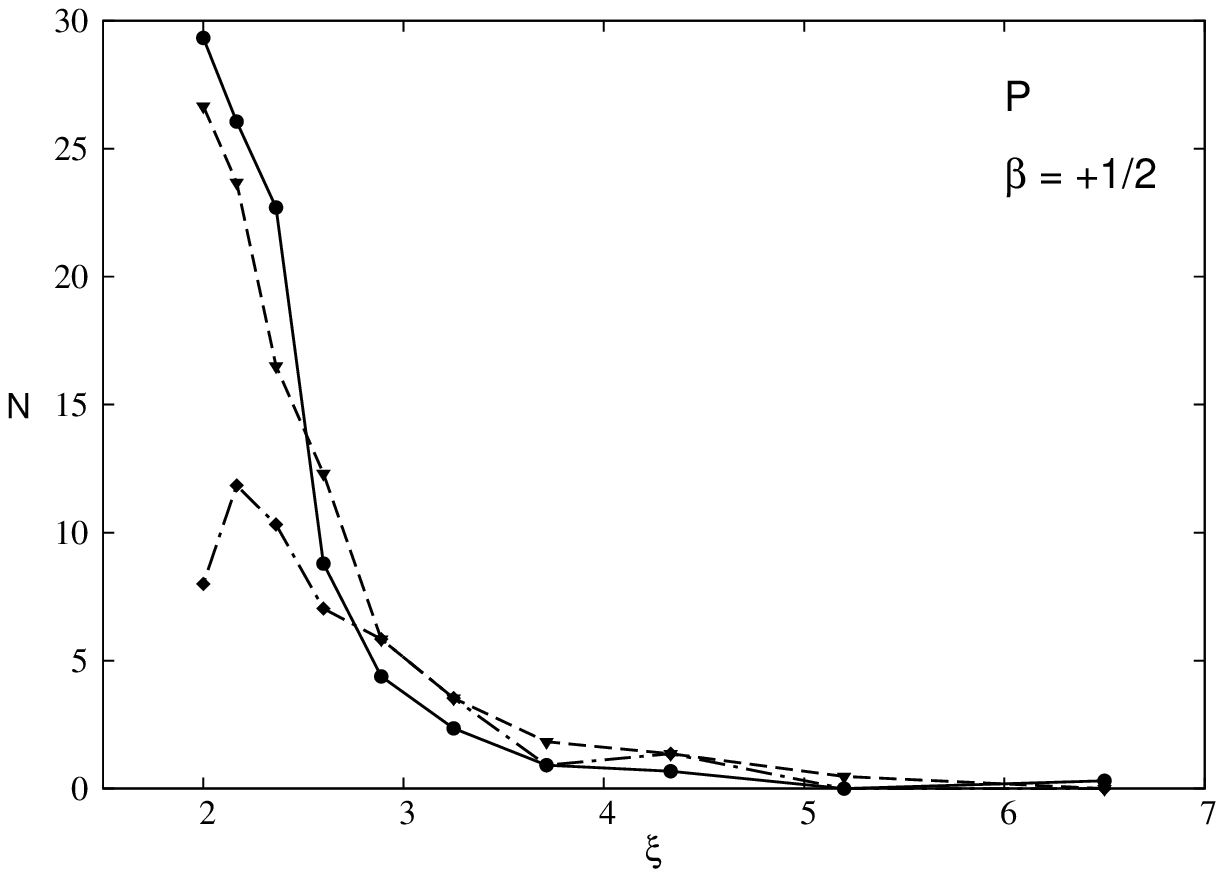}\label{fig:p_pl_m4}}
\end{minipage}
\caption{Number of stars ejected by the Globular Cluster with mass $M_{GC}=10^4$ M$_\odot$ as function of $\xi$, for Hernquist (left) and Plummer potential (right), for different values of the anisotropy parameter $\beta$. The different line styles refer to different GC orbit: $\alpha=0.1$ solid line, $\alpha=0.2$ dashed line, $\alpha=0.3$ dot-dashed line. $\alpha=0.4$ and $\alpha=0.5$ orbits give no ejected stars.}
\end{figure*}

\begin{figure*}
\centering
\begin{minipage}{20.5cm}
\subfigure{\includegraphics[scale=0.72]{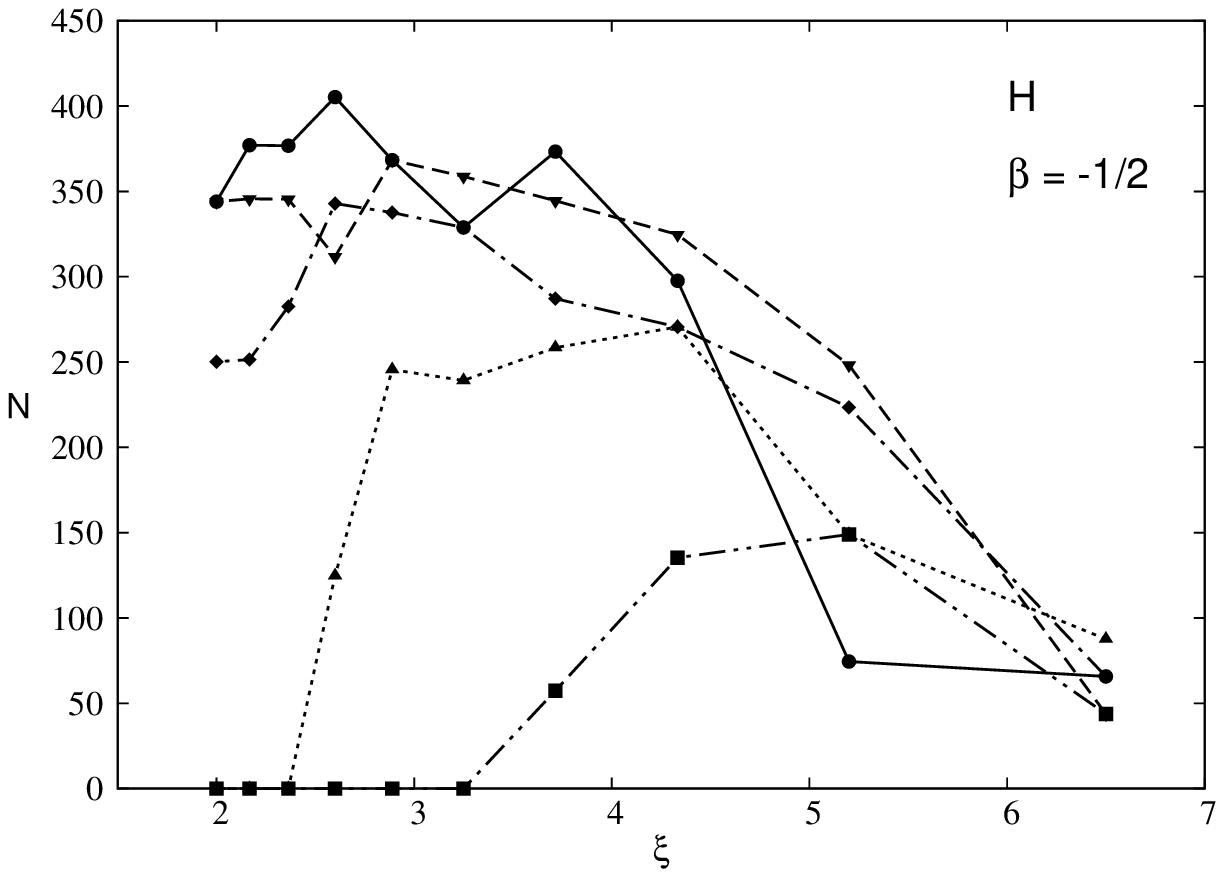}\label{fig:h_min_m5}}
\subfigure{\includegraphics[scale=0.72]{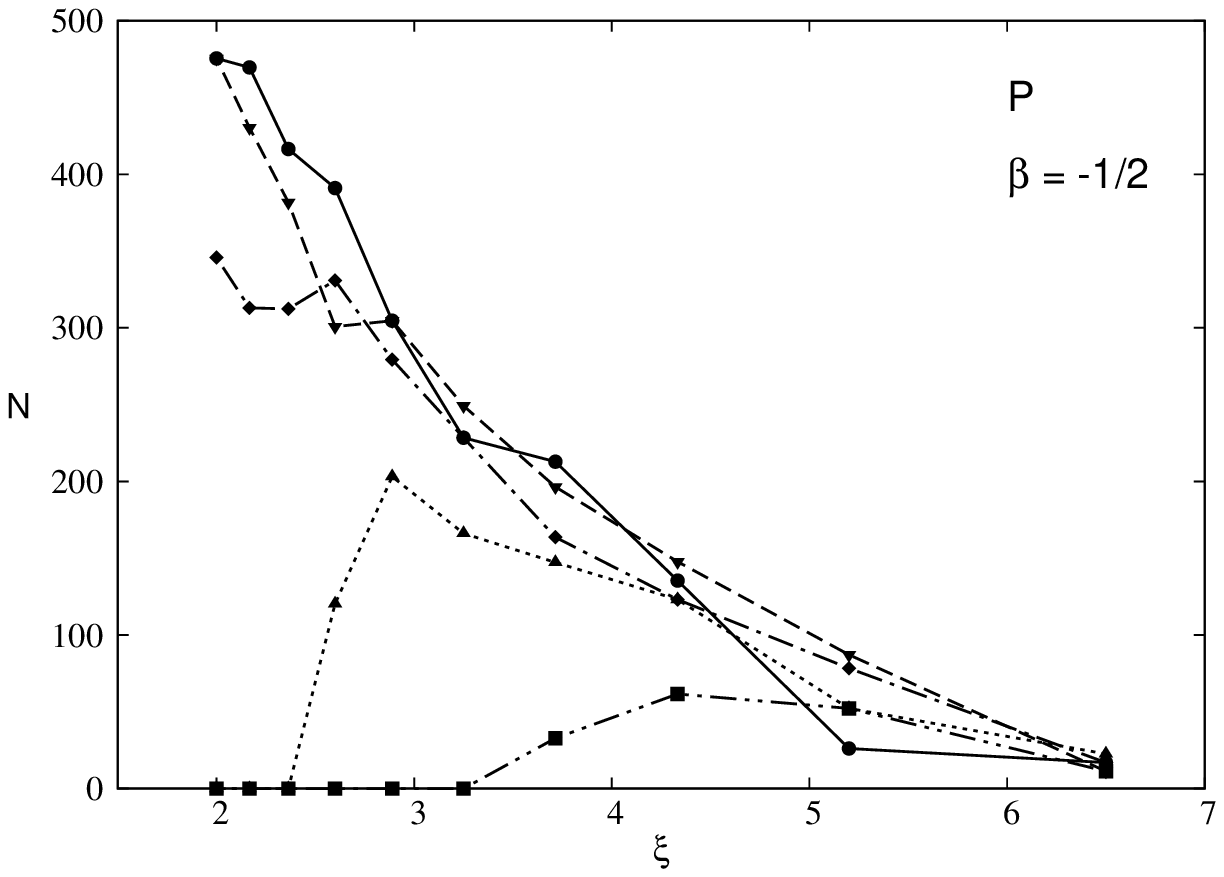}\label{fig:p_min_m5}}
\end{minipage}
\begin{minipage}{20.5cm}
\subfigure{\includegraphics[scale=0.72]{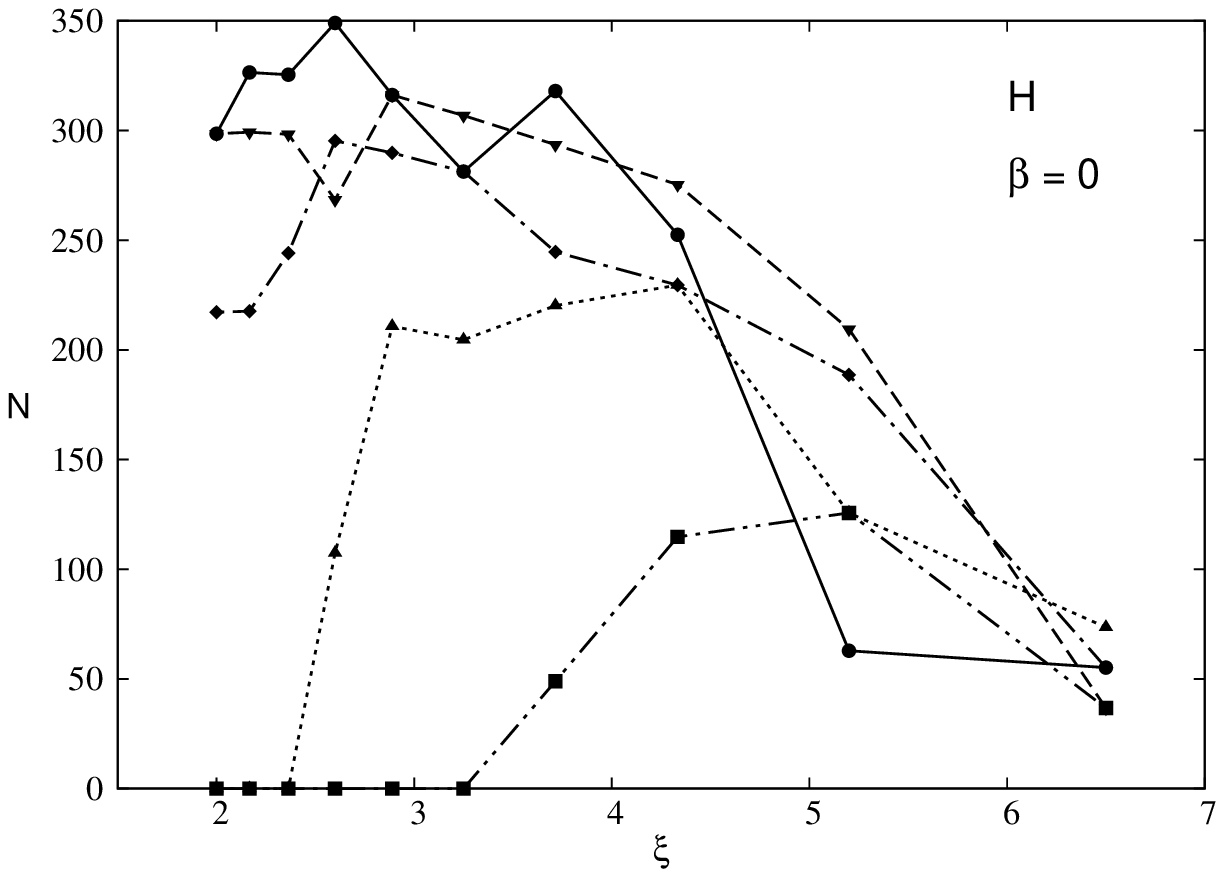}\label{fig:h_zer_m5}}
\subfigure{\includegraphics[scale=0.72]{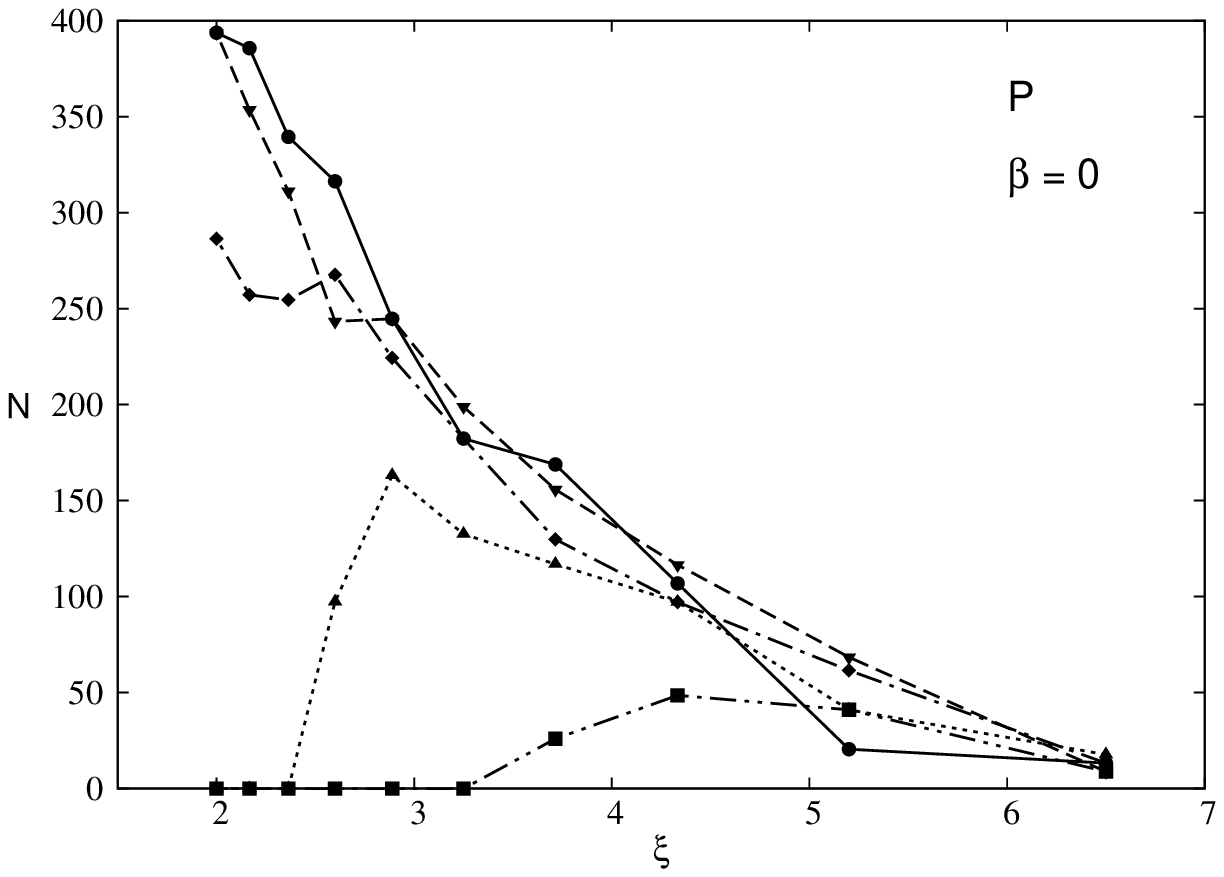}\label{fig:p_zer_m5}}
\end{minipage}
\begin{minipage}{20.5cm}
\subfigure{\includegraphics[scale=0.72]{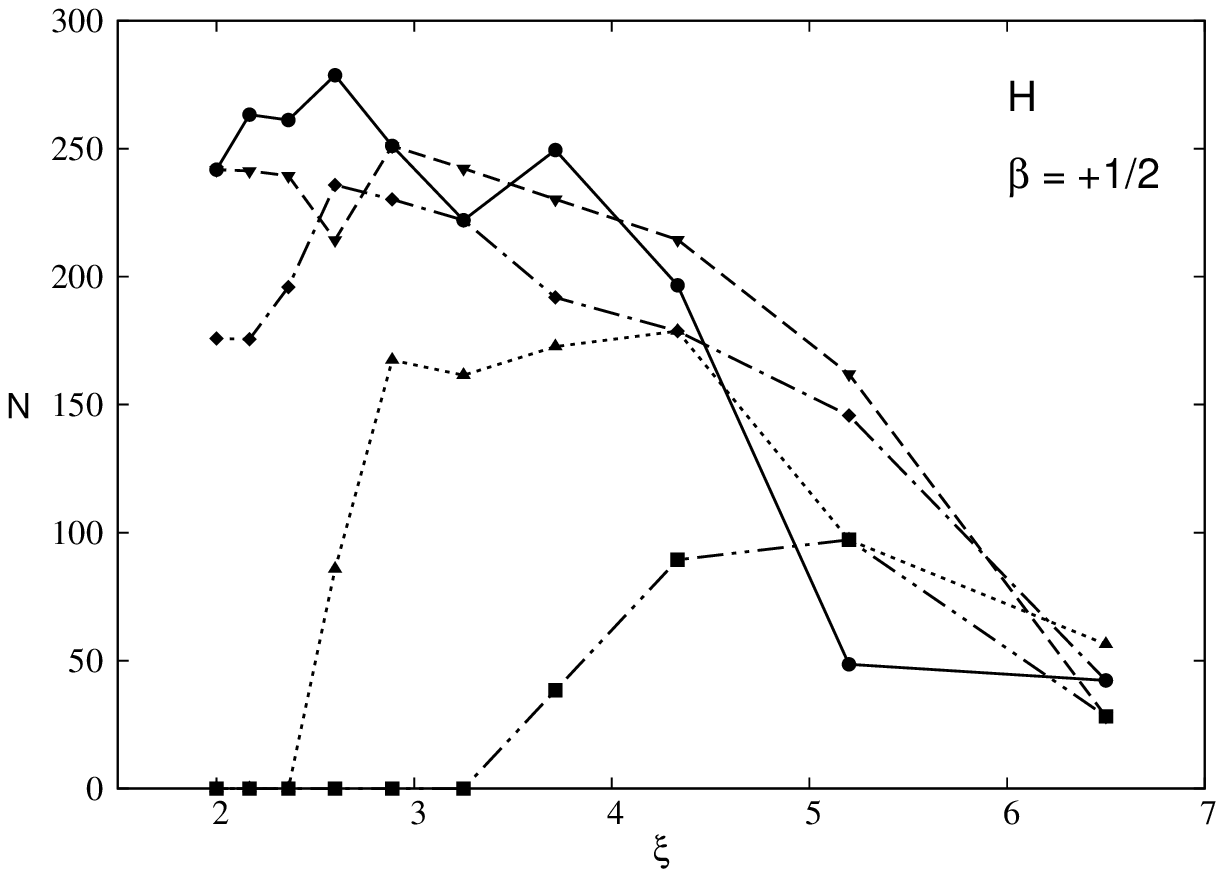}\label{fig:h_pl_m4}}
\subfigure{\includegraphics[scale=0.72]{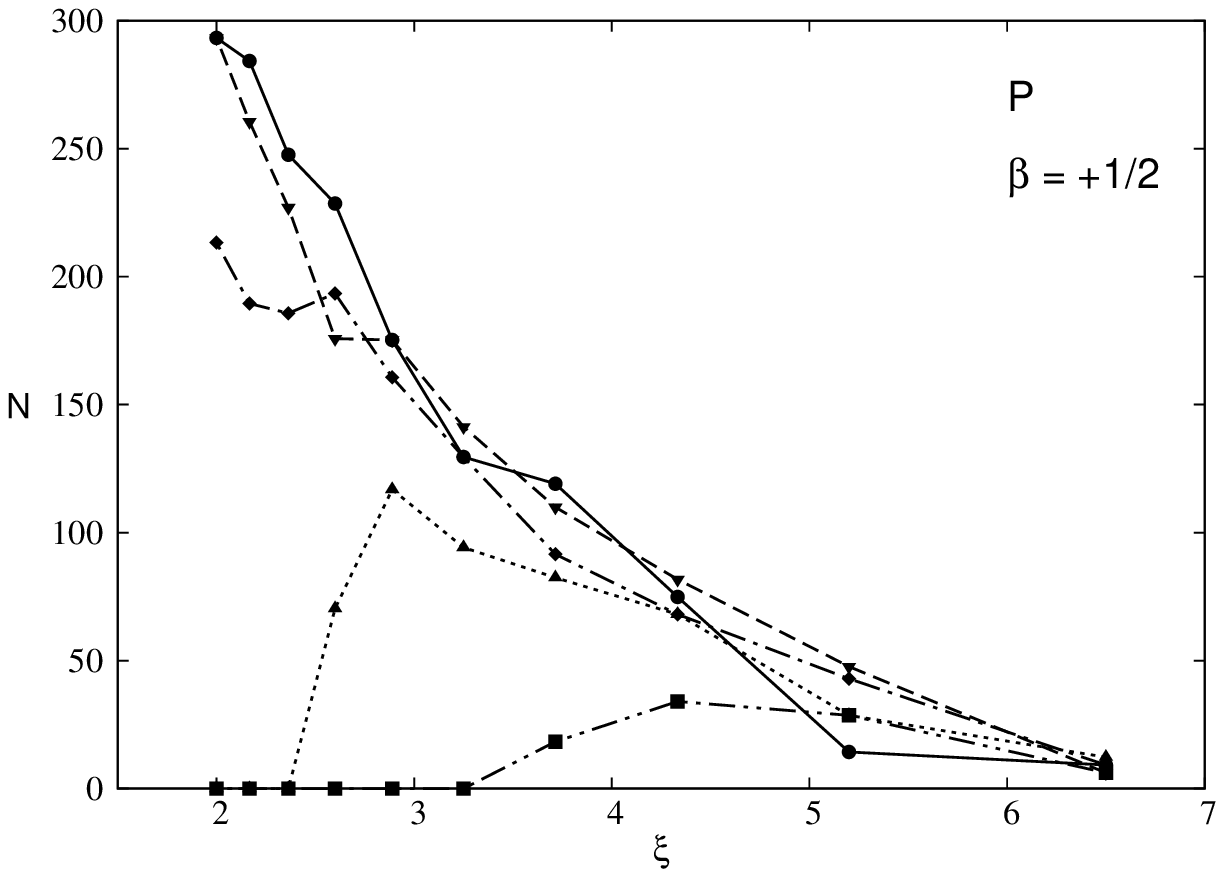}\label{fig:p_pl_m4}}
\end{minipage}
\caption{Number of stars ejected by the Globular Cluster with mass $M_{GC}=10^5$ M$_\odot$ as function of $\xi$, for Hernquist (left) and Plummer potential (right), for different values of the anisotropy parameter $\beta$. The different line styles refer to different GC orbit: $\alpha=0.1$ solid line, $\alpha=0.2$ dashed line, $\alpha=0.3$ dot-dashed line, $\alpha=0.4$ dotted line, $\alpha=0.5$ double dot-dashed line.}
\end{figure*}

\begin{figure*}
\centering
\begin{minipage}{20.5cm}
\subfigure{\includegraphics[scale=0.72]{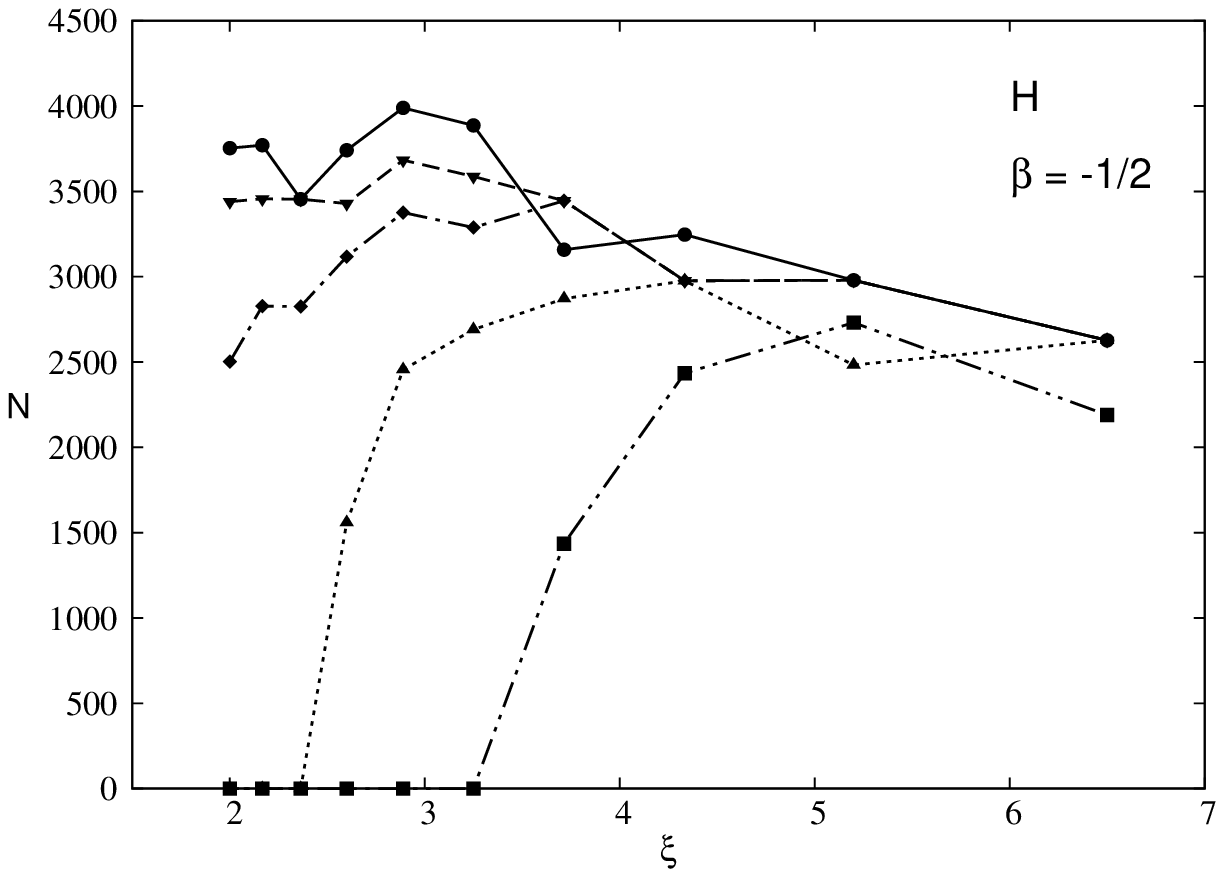}\label{fig:h_min_m6}}
\subfigure{\includegraphics[scale=0.72]{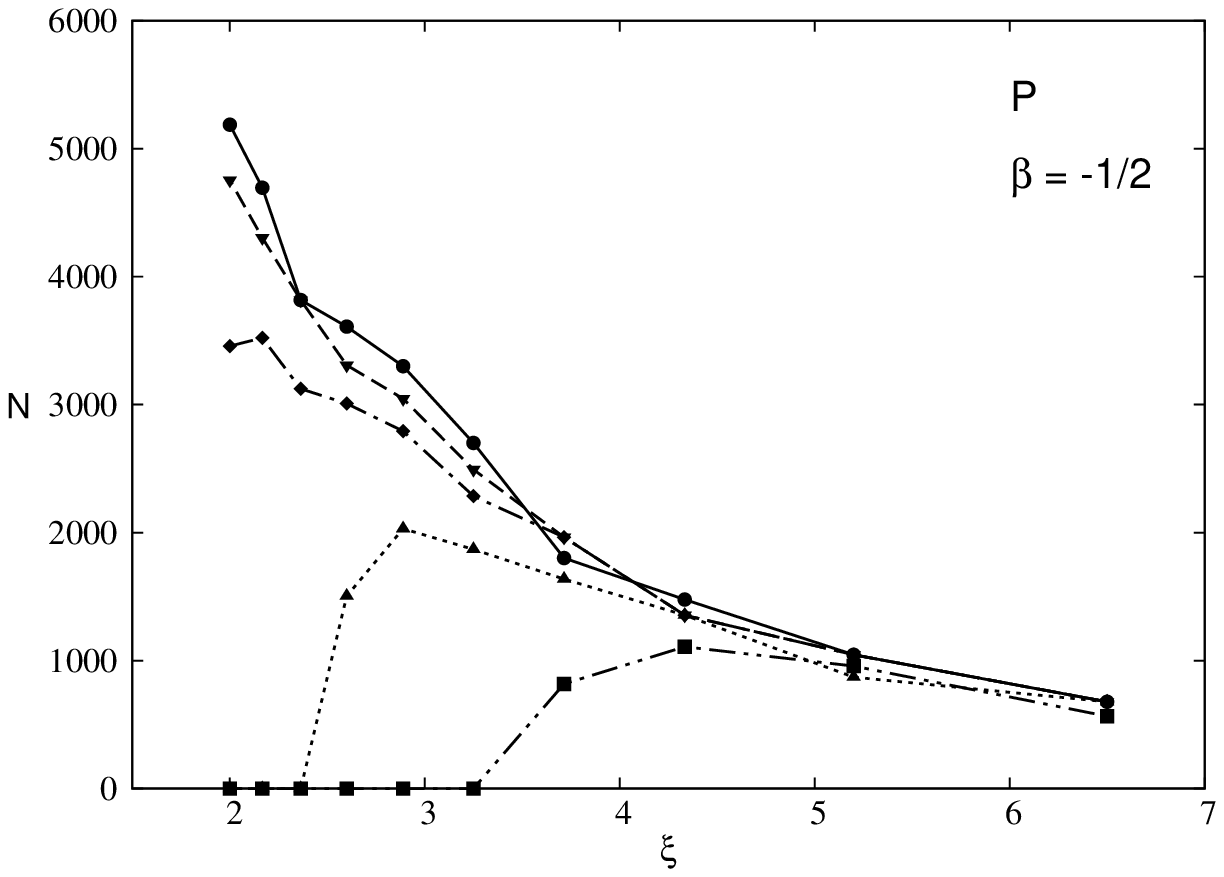}\label{fig:p_min_m6}}
\end{minipage}
\begin{minipage}{20.5cm}
\subfigure{\includegraphics[scale=0.72]{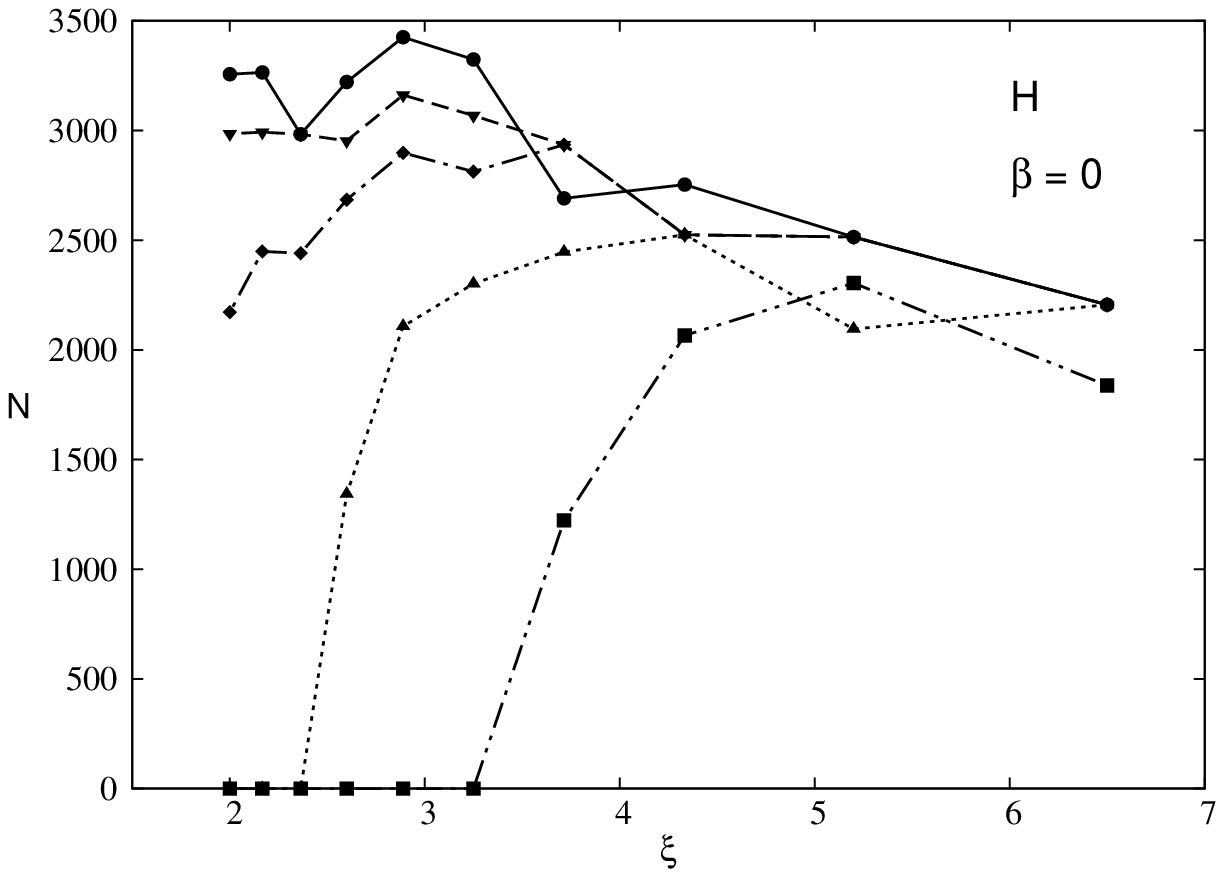}\label{fig:h_zer_m6}}
\subfigure{\includegraphics[scale=0.72]{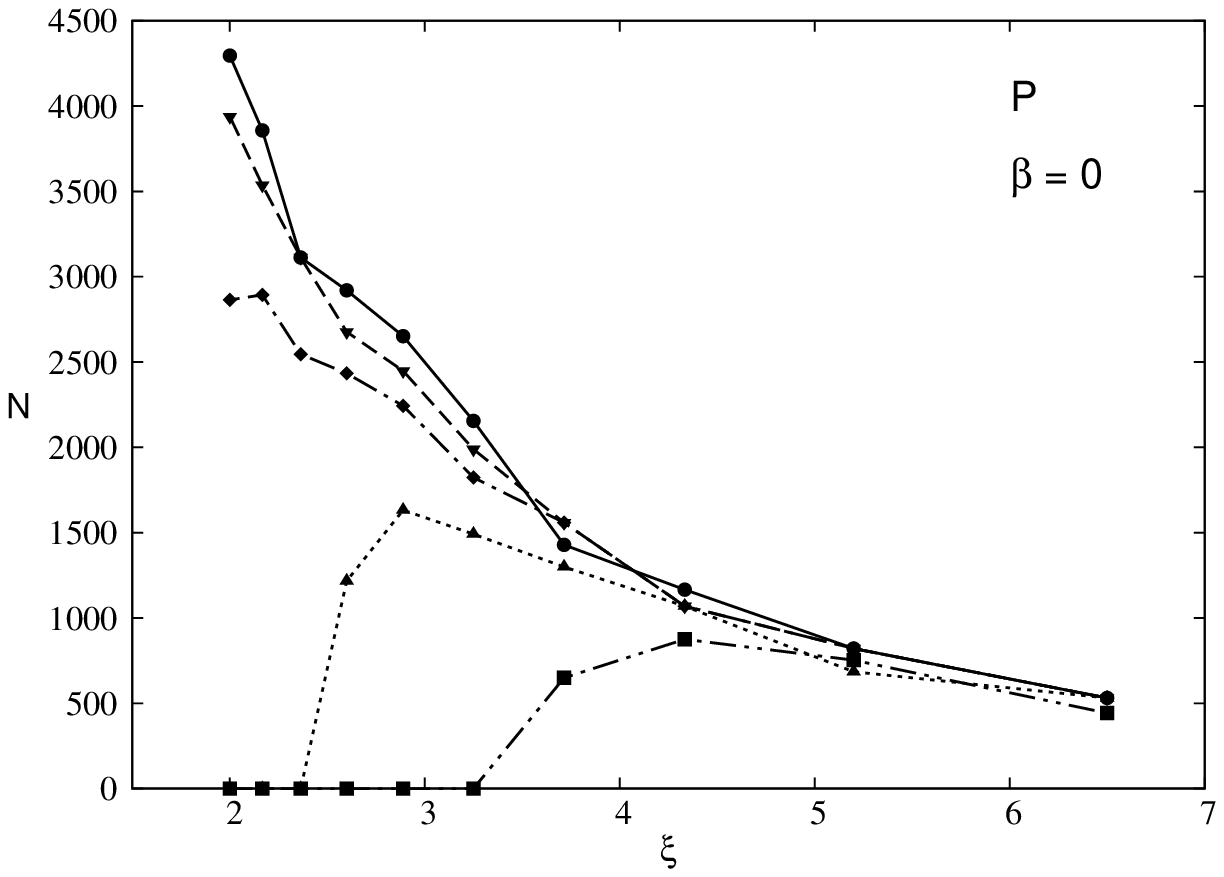}\label{fig:p_zer_m6}}
\end{minipage}
\begin{minipage}{20.5cm}
\subfigure{\includegraphics[scale=0.72]{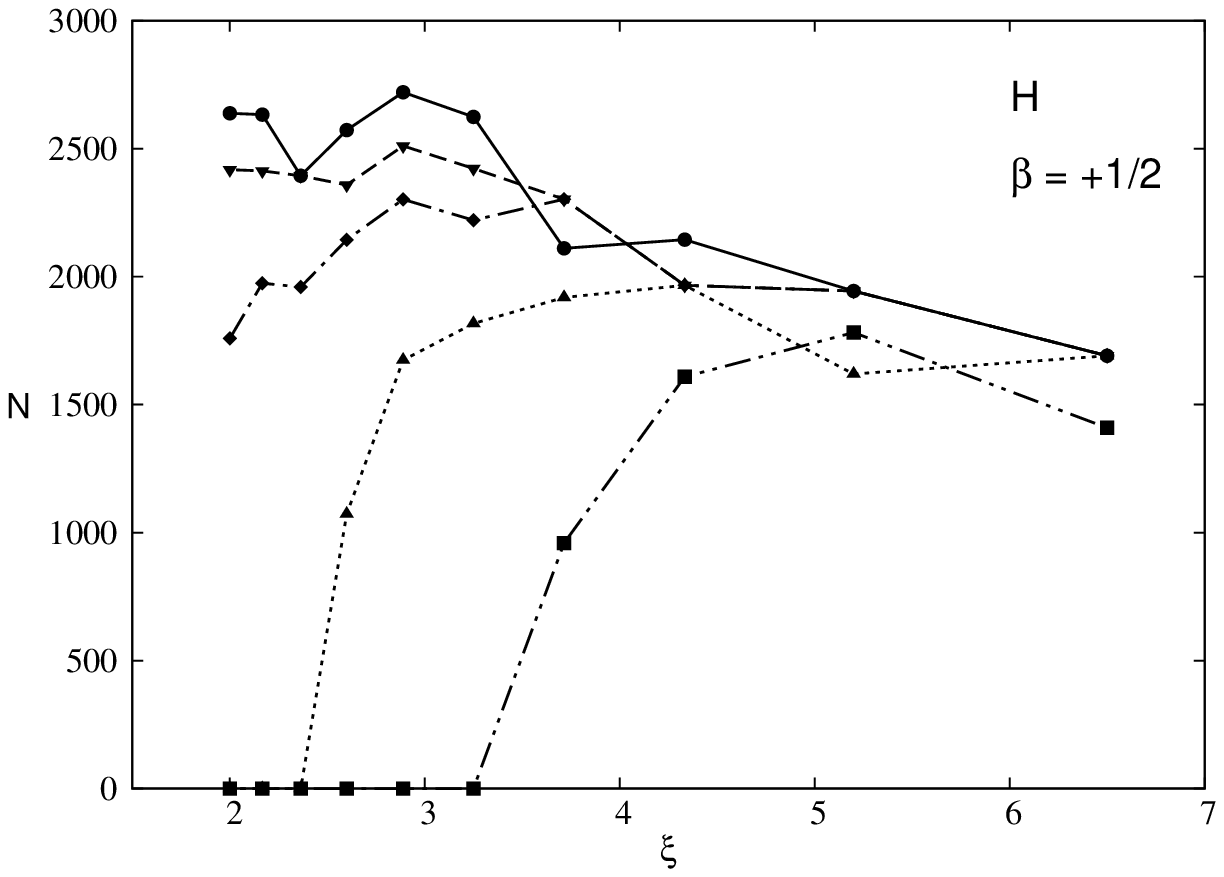}\label{fig:h_pl_m6}}
\subfigure{\includegraphics[scale=0.72]{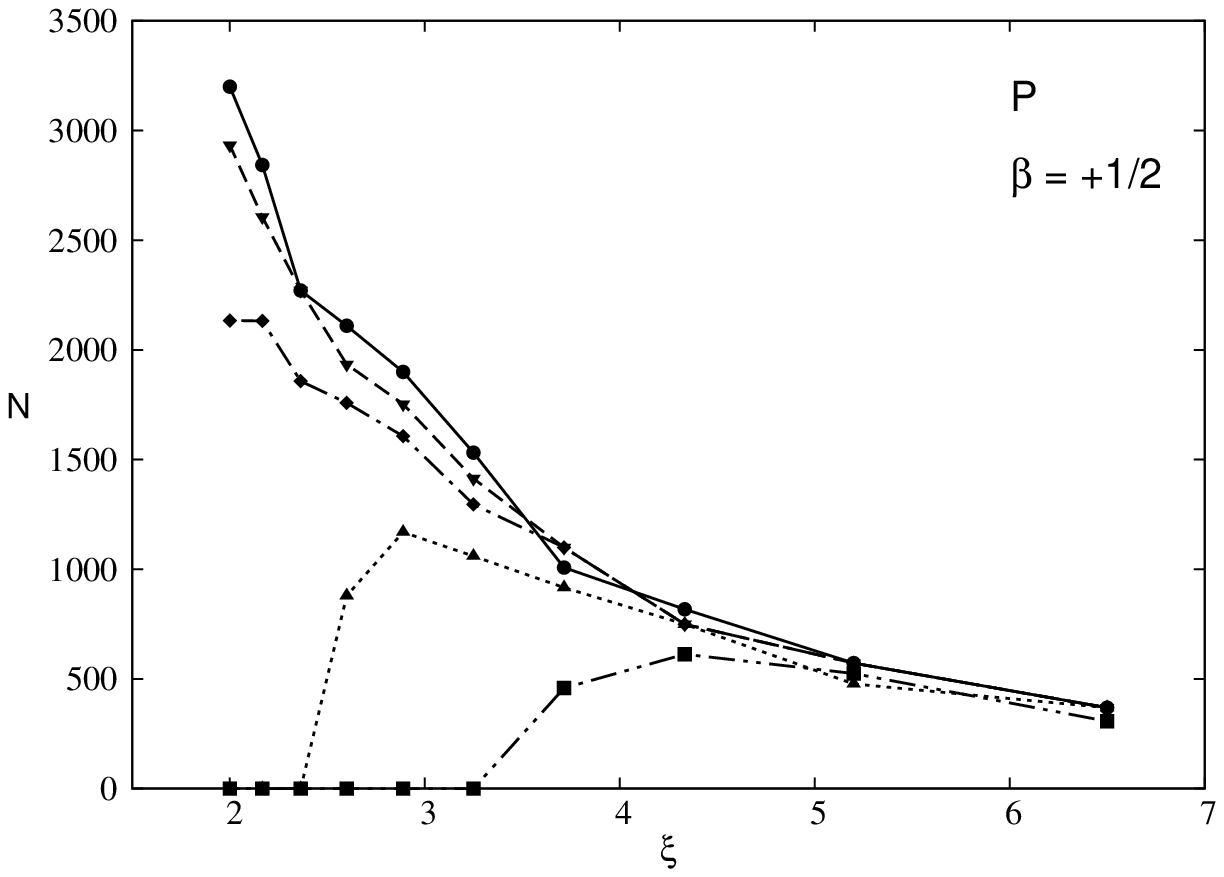}\label{fig:p_pl_m6}}
\end{minipage}
\caption{Number of stars ejected by the Globular Cluster with mass $M_{GC}=10^6$ M$_\odot$ as function of $\xi$, for Hernquist (left) and Plummer potential (right), for different values of the anisotropy parameter $\beta$. The different line styles refer to different GC orbit: $\alpha=0.1$ solid line, $\alpha=0.2$ dashed line, $\alpha=0.3$ dot-dashed line, $\alpha=0.4$ dotted line, $\alpha=0.5$ double dot-dashed line.}
\end{figure*}

Combining these evaluations of the local fractional abundance of stars on nearly circular orbits with the branching ratios  obtained in the previous sections, we can evaluate the actual numbers of stars which escape from or remain bound to the GC or the SMBH after close GC-SMBH interactions. 
This requires some assumptions. First of all we assume that the GC is composed by a single-mass ($m_{*}=1$ M$_{\odot}$) population of stars. Moreover, we assume that the core radius has a size which is half the innermost circular orbit radius chosen.

The number of stars in nearly circular orbits is evaluated by mean of the integral

\begin{equation} \label{eqn:num_c}
N_c=4\pi \mathcal{N}\int_{r_a}^{r_b} \nu_c(r)r^2 dr,
\end{equation}

where $\mathcal{N}=10^4$, $10^5$, $10^6$ is the assumed number of stars in the GCs, while $r_a$ and $r_b$ are given by

\begin{equation}
r_{a,b}=\frac{GM(r)}{v_{a,b}^2},
\end{equation}

being $v_{a,b}=v_c\pm\delta v_c$. A straightforward product of $N_c$ with the fractions of bound to the GC or BH or unbound (high velocity or hypervelocity) stars, as derived by our scattering experiments presented in the previous sections, gives an estimate of the actual number of stars ejected thanks to the GC-BH interaction mechanism. 

Figures 7-8-9 report the number of stars ejected from the GC after the close interaction between the GC and the BH, as function of the circular orbit radius for various values of $\alpha$, both in the isotropic ($\beta=0$) and anisotropic ($\beta=\pm 1/2$) Hernquist and Plummer GC models. The number of ejected stars depends, besides on the GC mass, on the GC potential model and on the degree of anisotropy, as shown by Fig. 6.

Moreover, the number of ejected stars increases with the GC mass. Actually this mechanism would produce $10\div 10^2$, $10^2\div 10^3$ and $10^3\div 10^4$ high velocity stars for GC mass of $10^4\ M_{\odot}$, $10^5\ M_{\odot}$, $10^6\ M_{\odot}$, respectively.

This almost linear dependence of the number of ejected stars on the GC mass is explained by that the gravitational energy per unit mass available to be converted into the test star kinetic energy is linearly scaling with $M_{GC}$, $E_{gr}\sim$ $M_{GC}$.

The same token holds for an explanation of the isotropic-anisotropic models difference. In the case of tangentially biased models ($\beta=-1/2$) there is a larger fraction of ejected stars because of the larger fraction of stars in nearly circular orbits. On the other hand, the fraction of ejected stars is smaller for radially biased models ($\beta=+1/2$).

\section{Conclusions}
\label{sec:con}
The phenomenon of the existence of high velocity stars, or even hypervelocity stars, in our Galaxy has been explained in the literature \citep{yut03} in terms of the presence of a massive black hole in the galactic centre ($\sim 4\times 10^6$ M$_\odot$ in our Milky Way). The mechanism of star acceleration to high speed requires, indeed, an efficient energy exchange in a multiple system, i.e. a three or more bodies interaction with the massive black hole \citep{hil88} or, if it exists, with a black hole binary \citep{ses06}.

In this paper we deepened what has been recently found by Arca Sedda et al. (in preparation) and preliminarily presented in \citep{sp15}, i.e. that the close passage of a massive globular cluster near a massive black hole can be source of ejection of stars from the cluster, which are accelerated to high speeds. In our study, we assumed $M_{BH}=10^8$ M$_\odot$ with the scope of identify at better the actual physical mechanism. The underlying mechanism is likely a 3-body interaction, where the ~\lq bodies\rq~ are the super massive black hole ($10^8$ M$_\odot$), the globular cluster ($10^4$, $10^5$ and $10^6$ M$_\odot$) and the test star (1 M$_\odot$) belonging to the globular cluster.

We have performed a series of high precision integration of the orbits of such 3 bodies to check the probability for the test star orbiting a globular cluster, which experiences a close (pericenter distance less than 10 pc) encounter, to remain bound to the cluster, to be captured by the black hole or gain high velocity such to overcome the cluster escape velocity and, possibly, the galaxy escape velocity. We determined the  branching ratios of these 3 phenomena and found that:
\begin{itemize}
\item 
the efficiency of the star acceleration process is almost linear in $M_{GC}$;
\item 
a massive globular cluster (composed by $10^6$ identical 1 M$_\odot$ stars) releases, in a single close passage around the super massive black hole, about $10^4$ stars;
\item 
in a very close GC-BH encounter ($M_{GC}=10^6$ M$_\odot$, $\alpha=0.1$) the probability of stars to remain bound, become bound to the BH or escape from the cluster are $\sim 5\%$, $\sim 45\%$, $\sim 50\%$, respectively;
\item
the fractions of stars, respect to the total ejected stars, which escape from the whole galaxy is $\sim 18\%$  for an $M_{tot}=7.81 \times 10^{10}$ M$_{\odot}$ elliptical and $\sim 0.5\%$  for an $M_{tot}=6.60 \times 10^{11}$ M$_{\odot}$ spiral galaxy.
\end{itemize}

Moreover, we studied the effects of the inclination of the star initial orbit around the GC respect to the GC orbit. When the orbits are perpendicular, the branching ratios of the ejected stars and of the stars captured by the BH decrease, while the branching ratio of stars which remain bound to the GC increases. Therefore, the overall effect of the increasing inclination of the star orbit respect to the GC orbit consists mainly in a slight reduction of the fraction of stars ejected from the GC and/or captured by the BH. Correspondingly, the velocity profile of the ejected stars is peaked at lower velcoity values.

Furthermore, we performed the same set of 3-body scattering experiments for the $M_{GC}=10^6$ GC assuming it is a Plummer sphere with a core radius of $0.5$ pc. In this smoothed case, the velocity distribution shrinks towards lower values of the velocity since the amount of gravitational energy, to convert into kinetic one, decreases. As a consequence, we found no ejection of hypervelocity stars, suggesting that, when the GC is described by a smoothed potential, it is likely to eject only high velocity stars, which escape the GC-BH system, but still remain bound to the host galaxy.

Different high velocity and hypervelocity ejection mechanisms predict different spatial distributions, velocity distributions and physical and kinematic characteristics of the ejected stars \citep{brw15}.

One important feature of this mechanism is the collimation of the ejected stars. We found that the high velocity stars are ejected in sort of jets, whose angular amplitude depends on the GC mass. High velocity stars produced by the interaction of a low-mass GC with a massive BH are likely to be concentrated in a small amplitude jet. On the other hand, higher GC masses make the ejection jets have a huge amplitude and stars are ejected in a nearly isotropic emission. These results are compatible with \citet{ses06} finding. Note that our results refer to a single GC-BH interaction; anyway, we expect a production of a high velocity star jet every time a GC undergoes a close encounter with a massive BH. Consequently, small clusters of high velocity stars are expected to be present in the sky, whose characteristics depend strongly on the mass of the GC from which they come from. At this regard, it is interesting to note that about half of the discovered HVS are clumped around the Leo constellation \citep{brw14}. The kinematics and dynamics of the jets, their position in the sky and their velocity profile may give information about the GC-BH interaction that produced them, in particular about the GC mass and its orbits before being disrupted by the massive black hole. Moreover, some stars of the jet, the ones on the right tail of  the velocity distribution, may have such high velocities to be HVS, and so to be lost by the jet itself since they are unbound respect to the hosting galaxy gravitational field. Being composed by stars formerly belonging to a GC, we expect that the jets contains stars of about same age and metallicity, with a common flight time. Hence, in principle, by measuring the dynamical and physical properties of such high velocity jets, it would be possible to infer information about the globular cluster progenitor before its tidal erosion by the  massive BH.

Finally, a counterpart of such jets is the production of a population of stars orbiting the inner galactic regions and which share age and metallicity with the jet stars.

As mentioned above, we assumed $M_{BH}=10^8$ M$_\odot$ with the scope of identify at better the underlying physical mechanism. For what concerns the Milky Way, we expect that the velocity distribution would be peaked at lower velocities due to the less energetic mechanism (being the SgrA* mass a factor 25 lower), with the production of  high velocity rather than hypervelocity stars. However, we suggest that, in the very last, eccentric and narrow GC orbits, some HVS may be produced through the mechanism studied in this work. Finally, we suggest that the recent observed runaway RR Lyrae variable star, MACHO 176.18833.411, \citep{kun15} may have been produced through the mechanism studied in this work.

\section*{Acknowledgments}

We thank S. Mikkola for making available to us his ARW code and for useful discussions about his use. We also thank E.M. Rossi for useful discussions about HVS and the mechanism studied in this work. Finally, we thank the anonymous referee for helpful comments and suggestions on this manuscript.

\appendix

\section{Fraction of particles in nearly circular orbits} \label{app:circ}

A spherical stellar system confined by a steady spherical potential $\Phi(r)$ can be described by a unique ergodic distribution function (DF). The DF depends on the phase-space coordinates only through the Hamiltonian $H(\textbf{r},\textbf{v})=\frac{1}{2}v^2+\Phi({\textbf r})$ \citep{bin11} and can be written as a non negative $f(\mathcal{E})$, where  $\mathcal{E}=-H+\Phi_0$ is the {\it relative} energy, with $\Phi_0$ a constant chosen such that $f>0$ for $\mathcal{E}>0$ and $f=0$ for $\mathcal{E}\le 0$.

The functional expression of $f(\mathcal{E})$, which obviously depends on the functional form of $\Phi(r)$, can be found by the Eddington's inversion formula

\begin{equation} \label{eqn:df_z}
f(\mathcal{E})=\frac{1}{\sqrt{8}\pi^2}\left[\int_{0}^{\mathcal{E}} \frac{d\Psi}{\sqrt{\mathcal{E}-\Psi}}\frac{d^2\nu}{d\Psi^2}+\frac{1}{\sqrt{\mathcal{E}}}\left.\left(\frac{d\nu}{d\Psi}\right)\right|_{\Psi=0}\right],
\end{equation}

where $\nu(\textbf{r})$ is the spatial number density generated by the DF

\begin{equation}
\nu(\textbf{r})=\int d^3 \textbf{v} f(\textbf{r},\textbf{v})=4\pi\int_0^{\sqrt{2\Psi}} dv\ v^2 f(\Psi-\frac{1}{2}v^2),
\end{equation}

and where $\Psi=-\Phi+\Phi_0$.

The Hernquist potential \citep{her90}, for which $\Phi_0=\Phi(r\rightarrow\infty)=0$,

\begin{equation} \label{eqn:hern}
\Phi(r)=-\frac{GM}{r+a},
\end{equation}

where $M$ is the total mass and $a$ is the core radius, leads, by Eq. \ref{eqn:df_z}, to

\begin{eqnarray}
f(\mathcal{E})&=&\frac{1}{\sqrt{2}(2\pi)^3(GMa)^(3/2)}\frac{\sqrt{\widetilde{\mathcal{E}}}}{(1-\widetilde{\mathcal{E}})^2}\times\nonumber\\
&\times& \left[(1-2\widetilde{\mathcal{E}})(8\widetilde{\mathcal{E}}^2-8\widetilde{\mathcal{E}}-3)+\frac{3\arcsin\sqrt{\widetilde{\mathcal{E}}}}{\sqrt{\widetilde{\mathcal{E}}(1-\widetilde{\mathcal{E}})}}\right],
\end{eqnarray}

where $\widetilde{\mathcal{E}}=\mathcal{E}/(GM/a)$ is the adimensional relative energy. The Plummer potential \citep{plu11}, for which $\Phi_0=\Phi(r\rightarrow\infty)=0$,

\begin{equation} \label{eqn:plumm}
\Phi(r)=-\frac{GM}{\sqrt{r^2+a^2}},
\end{equation}

where, again, $M$ is the total mass and $a$ is the core radius, leads, by Eq. \ref{eqn:df_z}, to

\begin{equation}
f(\mathcal{E})=\frac{96\widetilde{\mathcal{E}}^{7/2}}{7\sqrt{8}\pi^3(GMa)^(3/2)},
\end{equation}

where $\widetilde{\mathcal{E}}$ is, again, the adimensional relative energy.

The above DFs are isotropic, i.e. the anisotropy parameter

\begin{equation} \label{eqn:anis_p}
\beta=1-\frac{{\sigma}^2_{\theta}+{\sigma}^2_{\phi}}{2{\sigma}^2_r},
\end{equation}

is null. In Eq. \ref{eqn:anis_p}, $\sigma_r$, $\sigma_{\theta}$, $\sigma_{\phi}$ are the velocity dispersions on the $r$, $\theta$ and $\phi$ component, respectively, in a spherical polar reference frame.

Models with $\beta\ne 0$ can be generated by taking the distribution function in the form \citep{bin11}

\begin{equation} 
f(\mathcal{E},L)=L^{-2\beta}f_1(\mathcal{E}).
\label{noniso}
\end{equation}

In Eq. \ref{noniso}, $L$ is the absolute value of the specific angular momentum and $f_1(\mathcal{E})$ is an arbitrary non-negative function of $\mathcal{E}$. 
Given the DF in the form of Eq. \ref{noniso}, the space number density is given by 

\begin{eqnarray}
\nu(\textbf{r})&=&\int d^3 \textbf{v} f(\textbf{r},\textbf{v})= \\
&=&2\pi\int_0^{\pi} d\eta\ \sin\eta\int_0^{\sqrt{2\Psi}} dv\ v^2 f(\Psi-\frac{1}{2}v^2,rv \sin\eta). \nonumber
\end{eqnarray}

The expression of $f_1(\mathcal{E})$ depends on the form of the potential $\Phi(r)$ and on the value of the anisotropy parameter $\beta$. In the case $\beta=+1/2$ the expression of $f_1(\mathcal{E})$ is given by

\begin{equation}
f_1(\mathcal{E})=\frac{1}{2\pi^2}\left.\frac{d}{d\Psi}(r\nu)\right|_{\Psi=\mathcal{E}},
\end{equation}

while in the case $\beta=-1/2$ by

\begin{equation}
f_1(\mathcal{E})=\frac{1}{2\pi^2}\left.\frac{d^2}{d\Psi^2}(\nu/r)\right|_{\Psi=\mathcal{E}}.
\end{equation}
In the case of the Hernquist potential (\ref{eqn:hern}) we thus have  

\begin{equation}
f_1(\widetilde{\mathcal{E}})=\frac{3\widetilde{\mathcal{E}}}{4\pi^3 GMa},
\end{equation}

for $\beta=+1/2$, and

\begin{equation}
f_1(\widetilde{\mathcal{E}})=\frac{1}{2\pi^3\left(GMa\right)^2}\frac{\widetilde{\mathcal{E}}^5-10\widetilde{\mathcal{E}}^4+10\widetilde{\mathcal{E}}^3}{1-\widetilde{\mathcal{E}}^4},
\end{equation}

for $\beta=-1/2$.\\
The Plummer potential (\ref{eqn:plumm}), instead, leads to

\begin{equation}
f_1(\widetilde{\mathcal{E}})=\frac{3}{8\pi^3 GMb}\frac{4\widetilde{\mathcal{E}}^3-5\widetilde{\mathcal{E}}^5}{(1-\widetilde{\mathcal{E}}^2)^{1/2}},
\end{equation}

for $\beta=+1/2$, and to 

\begin{equation}
f_1(\widetilde{\mathcal{E}})=\frac{3}{8\pi^3 (GMb)^2}\frac{30\widetilde{\mathcal{E}}^4-47\widetilde{\mathcal{E}}^6+20\widetilde{\mathcal{E}}^8}{(1-\widetilde{\mathcal{E}}^2)^{1/2}},
\end{equation}

for $\beta=-1/2$.

Finally, to calculate the fraction of stars on nearly circular orbits $G(\xi)=\nu_c(\xi)/\nu(\xi)$, we consider a~\lq tolerance\rq~$\delta$, which quantifies the departure from the exact circular velocity. In our calculations we consider in $\nu_c(\xi)$ all the stars having a local speed in the interval $-\delta \leq v/v_c \leq +\delta$. Conseqeuently, their number density is given by 

\begin{eqnarray} \label{eqn:nucir}
\nu_c(\textbf{r})&=&\int d^3 \textbf{v} f(\textbf{r},\textbf{v})= \\
&=&I_{\beta}\int_{v_c-\delta v_c}^{v_c+\delta v_c} dv\ v^2 f(\Psi-\frac{1}{2}v^2,rv \sin\eta), \nonumber
\end{eqnarray}

where

\begin{equation}
I_{\beta}= \begin{cases} 4\pi & \beta=0 \\
2\pi\int_0^{\pi} d\eta\ \sin\eta & \beta=\pm 1/2. \end{cases}
\end{equation}

\bsp

\label{lastpage}


\begin{thebibliography}{99}
\bibitem[\protect\citeauthoryear{Antonini et al.}{2012}]{ant12} Antonini F., Capuzzo-Dolcetta R., Mastrobuono-Battisti A., Merritt D., 2012, ApJ, 750, 111
\bibitem[\protect\citeauthoryear{Baumgardt, Gualandris \& Portegies Zwart}{Baumgardt et al.}{2006}]{bau06} Baumgardt H., Gualandris A., Portegies Zwart S.F., 2006, J. Phys.: Conf. Ser., 54.1, 301
\bibitem[\protect\citeauthoryear{Binney}{1981}]{bin81} Binney J., 1981, MNRAS, 196, 455
\bibitem[\protect\citeauthoryear{Binney \& Tremaine}{2011}]{bin11} Binney J., Tremaine S., 2011, Galactic Dynamics. Princeton university press, Princeton
\bibitem[\protect\citeauthoryear{Blaauw}{1961}]{bla61} Blaauw A., 1961, Bull. Astron. Inst. Netherlands, 15, 265
\bibitem[\protect\citeauthoryear{Bonanos et al.}{2008}]{bon08} Bonanos A.Z., Lopez-Morales M., Hunter I., Ryans R.S.I., 2008, ApJ, 675, L77
\bibitem[\protect\citeauthoryear{Bromley et al.}{2006}]{brm06} Bromley B.C., Kenyon S.J., Geller M.J., Barcikowski E., Brown W.R., Kurtz M.J., 2006, ApJ, 653, 1194
\bibitem[\protect\citeauthoryear{Brown}{2015}]{brw15} Brown W.R., 2015, Annu. Rev. Astron. Astrophys., 53, 15
\bibitem[\protect\citeauthoryear{Brown, Geller \& Kenyon}{Brown et al.}{2009}]{brw09} Brown W.R., Geller M.J., Kenyon S.J., 2009, ApJ, 690.2, 1639
\bibitem[\protect\citeauthoryear{Brown, Geller \& Kenyon}{Brown et al.}{2014}]{brw14} Brown W.R., Geller M.J., Kenyon S.J., 2014, ApJ, 787, 89
\bibitem[\protect\citeauthoryear{Brown et al.}{2005}]{brw05} Brown W.R., Geller M.J., Kenyon S.J., Kurtz M.J., 2005, ApJ Lett., 622, L33
\bibitem[\protect\citeauthoryear{Brown et al.}{2010}]{brw10} Brown W.R. et al., 2010, ApJ Lett., 719, L23
\bibitem[\protect\citeauthoryear{Bulirsch \& Stoer}{1966}]{bul66} Bulirsch R., Stoer J., 1966, Numer. Math., 8, 1
\bibitem[\protect\citeauthoryear{Capuzzo-Dolcetta}{1993}]{cap93} Capuzzo-Dolcetta R., 1993, ApJ, 415, 616
\bibitem[\protect\citeauthoryear{Capuzzo-Dolcetta \& Miocchi}{2008}]{cap08} Capuzzo-Dolcetta R., Miocchi P., MNRAS Lett., 388, L69
\bibitem[\protect\citeauthoryear{Fujita}{2009}]{fuj09} Fujita Y., 2009, ApJ, 691, 1050
\bibitem[\protect\citeauthoryear{Ginsburg \& Loeb}{2006}]{gin06} Ginsburg I., Loeb A., 2006, MNRAS, 368, 221
\bibitem[\protect\citeauthoryear{Ginsburg, Loeb, Wegner}{Ginsburg et al.}{2012}]{gin12} Ginsburg I., Loeb A., Wegner G.A., 2012, MNRAS, 423.1, 948
\bibitem[\protect\citeauthoryear{Gnedin et al.}{2005}]{gne05} Gnedin O.Y., Gould A., Miralda-Escudé J., Zentner A.R., 2005, ApJ, 634, 344.
\bibitem[\protect\citeauthoryear{Gould \& Quillen}{2003}]{gou03} Gould A., Quillen A.C., 2003, ApJ, 592, 935
\bibitem[\protect\citeauthoryear{Gualandris \& Portegies Zwart}{2007}]{gua07} Gualandris A., Portegies Zwart S.F., 2007, MNRAS Lett., 376.1, L29
\bibitem[\protect\citeauthoryear{Gualandris, Portegies Zwart \& Sipior}{Gualandris et al.}{2005}]{gua05} Gualandris A., Portegies Zwart S.F., Sipior M.S., 2005, MNRAS, 363, 223
\bibitem[\protect\citeauthoryear{Gvaramadze}{2009}]{gvr09} Gvaramadze V.V., 2009, MNRAS, 395, L85
\bibitem[\protect\citeauthoryear{Gvaramadze \& Gualandris}{2011}]{gva11} Gvaramadze V.V., Gualandris A., 2011, MNRAS, 410, 304
\bibitem[\protect\citeauthoryear{Gvaramadze, Gualandris \& Portegies Zwart}{Gvaramadze et al.}{2009}]{gva09} Gvaramadze V.V., Gualandris A., Portegies Zwart S.F., 2009, MNRAS, 396, 570
\bibitem[\protect\citeauthoryear{Hansen}{2007}]{han07} Hansen B.M.S., 2007, ApJ Lett., 671.2, L133
\bibitem[\protect\citeauthoryear{Hellstr\"{o}m \& Mikkola}{2010}]{hel10} Hellstr\"{o}m C., Mikkola S., 2010, Celest. Mech. Dyn. Astron., 106, 143
\bibitem[\protect\citeauthoryear{Hernquist}{1990}]{her90} Hernquist L., 1990, ApJ, 356, 359
\bibitem[\protect\citeauthoryear{Hills}{1988}]{hil88} Hills J.G., 1988, Nature, 331, 687
\bibitem[\protect\citeauthoryear{Hoogerwerf, de Bruijne \& de Zeeuw}{Hoogerwerf et al.}{2001}]{hoo01} Hoogerwerf R., de Bruijne J.H.J., de Zeeuw P.T., 2001, A \& A, 365, 49
\bibitem[\protect\citeauthoryear{Humason \& Zwicky}{1947}]{hum47} Humason M.L., Zwicky F., 1947, ApJ, 105, 85
\bibitem[\protect\citeauthoryear{Kobayashi et al.}{2012}]{kob12} Kobayashi S., Hainick Y., Sari R., Rossi E.M., 2012, ApJ, 670, 747
\bibitem[\protect\citeauthoryear{Kollmeier et al.}{2010}]{kol10} Kollmeier J.A. et al., 2010, ApJ, 723.1, 812
\bibitem[\protect\citeauthoryear{Kollmeier et al.}{2009}]{kol09} Kollmeier J.A., Gould A., Knapp G., Beers T.C., 2009, ApJ, 697, 1543
\bibitem[\protect\citeauthoryear{Kornreich \& Lovelace}{2008}]{krn08} Kornreich D.A., Lovelace R.V.E., 2008, ApJ, 681, 104
\bibitem[\protect\citeauthoryear{Kunder et al.}{2015}]{kun15} Kunder A. et al., 2015, ApJ Lett., 808, L12
\bibitem[\protect\citeauthoryear{Leonard \& Duncan}{1990}]{leo90} Leonard P.J.T., Duncan M.J., 1990, AJ, 99, 608
\bibitem[\protect\citeauthoryear{Li et al.}{2015}]{li15} Li Y. et al., 2015, preprint (arXiv:1506.01818v2)
\bibitem[\protect\citeauthoryear{Lingam}{2014}]{lin14} Lingam M., 2014, Astrophys. Space Sci., 354, 561
\bibitem[\protect\citeauthoryear{Loeb}{2011}]{loe11} Loeb A., 2011, J. Cosm. Astrop. Phys., 2011.04, 023
\bibitem[\protect\citeauthoryear{Lopez-Morales \& Bonanos}{2008}]{lop08} Lopez-Morales M., Bonanos A.Z., 2008, ApJ lett., 685, L47
\bibitem[\protect\citeauthoryear{Marconi \& Hunt}{2003}]{mar03} Marconi A., Hunt L.K., 2003 ApJ Lett. 589, L21
\bibitem[\protect\citeauthoryear{Mikkola \& Aarseth}{2001}]{mik01} Mikkola S., Aarseth S., 2001, Celest. Mech. Dyn. Astron., 84, 343
\bibitem[\protect\citeauthoryear{Mikkola \& Merritt}{2006}]{mik06} Mikkola S., Merritt D., 2006, MNRAS, 372, 219
\bibitem[\protect\citeauthoryear{Mikkola \& Merritt}{2008}]{mik08} Mikkola S., Merritt D., 2008, AJ, 135, 2398
\bibitem[\protect\citeauthoryear{Mikkola \& Tanikawa}{1999a}]{mik99a} Mikkola S., Tanikawa K., 1999a, Celest. Mech. Dyn. Astron., 74, 287
\bibitem[\protect\citeauthoryear{Mikkola \& Tanikawa}{1999b}]{mik99b} Mikkola S., Tanikawa K., 1999b, MNRAS, 310, 745
\bibitem[\protect\citeauthoryear{Miyamoto \& Nagai}{1975}]{miy75} Miyamoto M., Nagai R., 1975, Publ. Astron. Soc. Jpn., 27, 533
\bibitem[\protect\citeauthoryear{O'Leary \& Loeb}{2008}]{oll08} O'Leary R. M., Loeb A., 2008, MNRAS, 383, 86
\bibitem[\protect\citeauthoryear{Palladino et al.}{2014}]{pal14} Palladino L.E. et al., 2014, ApJ, 780, 7
\bibitem[\protect\citeauthoryear{Perets}{2009}]{per09} Perets H.B., 2009, ApJ, 698, 1330
\bibitem[\protect\citeauthoryear{Perets, Hopman \& Alexander}{Perets et al.}{2007}]{per07} Perets H.B., Hopman C., Alexander T., 2007, ApJ, 656, 709
\bibitem[\protect\citeauthoryear{Perets \& Subr}{2012}]{per12} Perets H.B., Subr L., 2012, ApJ, 751, 133
\bibitem[\protect\citeauthoryear{Plummer}{1911}]{plu11} Plummer H.C., 1911, MNRAS, 71, 140
\bibitem[\protect\citeauthoryear{Portegies Zwart}{2000}]{por00} Portegies Zwart S.F., 2000, ApJ, 544, 437
\bibitem[\protect\citeauthoryear{Poveda, Ruiz \& Allen}{Poveda et al.}{1967}]{pov67} Poveda A., Ruiz J., Allen C., 1967, Bol. Obser. Tonantzintla y Tacubaya, 4, 86
\bibitem[\protect\citeauthoryear{Przybilla et al.}{2008}]{prz08} Przybilla N., Nieva M.F., Heber U., Butler K., 2008, ApJ Lett., 684, L103
\bibitem[\protect\citeauthoryear{Rossi, Kobayashi \& Sari}{Rossi et al.}{2014}]{ros14} Rossi E.M., Kobayashi S., Sari R., 2014, ApJ, 795.2, 125
\bibitem[\protect\citeauthoryear{Rownd, Dickey \& Helou}{Rownd et al.}{1994}]{rwn94} Rownd B.K., Dickey J.M., Helou G. 1994, AJ, 108, 1638
\bibitem[\protect\citeauthoryear{Sari, Kobayashi \& Rossi}{Sari et al.}{2009}]{sar09} Sari R., Kobayashi S., Rossi E.M., 2010, ApJ, 708, 605
\bibitem[\protect\citeauthoryear{Scheck et al.}{2006}]{sch06} Scheck L., Kifonidis K., Janka H.-T., M\"{u}ller E., 2006, A \& A, 457, 963
\bibitem[\protect\citeauthoryear{Sesana, Haardt \& Madau}{Sesana et al.}{2006}]{ses06} Sesana A., Haardt F., Madau P., 2006, ApJ, 651.1, 392
\bibitem[\protect\citeauthoryear{Sesana, Haardt \& Madau}{Sesana et al.}{2007}]{ses07} Sesana A., Haardt F., Madau P., 2007, MNRAS Lett., 379, L45
\bibitem[\protect\citeauthoryear{Sherwin, Loeb \& O'Leary}{Sherwin et al.}{2008}]{she08} Sherwin B., Loeb A., O'Leary R., 2008, MNRAS, 386, 1179
\bibitem[\protect\citeauthoryear{Silva \& Napiwotzki}{2011}]{sil11} Silva M.D.V., Napiwotzki R., 2011, MNRAS, 411, 2596
\bibitem[\protect\citeauthoryear{Spera, Arca-Sedda, Capuzzo-Dolcetta}{Spera et al.}{2015}]{sp15} Spera M., Arca-Sedda M., Capuzzo-Dolcetta R., 2015,  Proc. of IAU Symp. 312, in press (arXiv:1501.03175) 
\bibitem[\protect\citeauthoryear{Szebehely}{1966}]{sze66} Szebehely V., 1966, Theory of orbits. Acad. Press, New York
\bibitem[\protect\citeauthoryear{Tutukov \& Federova}{2009}]{tuf09} Tutukov A.V., Fedorova A.V., 2009, Astron. Rep., 53.9, 839
\bibitem[\protect\citeauthoryear{Vickers, Smith \& Grebel}{Vickers et al.}{2015}]{vic15} Vickers J.J., Smith M.C., Grebel E.K., 2015, AJ, 150.3, 77
\bibitem[\protect\citeauthoryear{Wang, Sulkanen \& Lovelace}{Wang et al.}{1992}]{wan92} Wang J.C.L., Sulkanen M.E., Lovelace R.V.E., 1992, ApJ, 390, 46
\bibitem[\protect\citeauthoryear{Yu \& Madau}{2007}]{yum07} Yu Q., Madau P., 2007, MNRAS, 379, 1293
\bibitem[\protect\citeauthoryear{Yu \& Tremaine}{2003}]{yut03} Yu Q., Tremaine S., 2003, ApJ, 599, 1129
\bibitem[\protect\citeauthoryear{Zhong et al.}{2014}]{zho14} Zhong J. et al., 2014, ApJ, 789, L2
\bibitem[\protect\citeauthoryear{Zubovas, Wynn \& Gualandris}{2013}]{zub13} Zubovas K., Wynn G.A., Gualandris A., 2013, ApJ, 771.2, 118
\end{thebibliography}
\end{document}